\documentclass[aps, amssymb, amsmath, superscriptaddress, 
twocolumn] {revtex4}

\usepackage{chngcntr}

\usepackage{graphicx}
\usepackage{color}
\usepackage{amsmath}
\usepackage{enumitem}
\usepackage{amssymb}
\usepackage{hyperref}
\usepackage{cancel}
\usepackage{ulem}
\usepackage{multirow}

\newcommand{\be}{\begin{equation}}
\newcommand{\ee}{\end{equation}}

\newcommand{\bea}{\begin{eqnarray}}
\newcommand{\eea}{\end{eqnarray}}

\newcommand{\lp}{\left(}
\newcommand{\rp}{\right)}
\newcommand{\sgn}{{\rm sgn}}

\renewcommand{\vec}[1]{{\boldsymbol #1}}

\begin{document}
\title{Spin-triplet superconductivity at the onset of isospin order in biased bilayer graphene}

\author{Zhiyu Dong${}^1$, Andrey V. Chubukov${}^2$, Leonid Levitov} 
\affiliation{Department of Physics, Massachusetts Institute of Technology, Cambridge, MA 02139, USA\\ ${}^2$W. I. Fine Theoretical Physics Institute, University of Minnesota, Minneapolis, MN 55455, USA}

%
%

%
%

\begin{abstract}
The quest for unconventional superconductivity governed by Coulomb repulsion between electrons rather than phonon attraction received new momentum with the advent of moir\'e graphene. Initially, delineating the phonon and Coulomb-repulsion-based pairing mechanisms has proven to be a challenging task, however the situation has changed after recent discovery of superconductivity in non-twisted graphene bilayers and trilayers. Superconductivity occurring at the phase boundaries of spin and valley polarized orders calls for non-phonon scenarios, yet the specific pairing mechanisms remain to be understood. Here we analyze a striking example --- superconductivity in graphene bilayers occurring at the onset of valley-polarized order induced by a magnetic field. We describe an attraction-from-repulsion mechanism for pairing mediated by a quantum-critical mode, which fully explains the observed phenomenology. While it is usually notoriously difficult to infer the pairing mechanism from the observed superconducting phases, this case presents a rare exception, allowing for a fairly unambiguous identification of the origin of the pairing glue. A combination of factors such as the location of superconducting phase at the onset of isospin-polarized phase, a threshold in a magnetic field, above which superconductivity occurs, and its resilience at high magnetic fields paints a clear picture of a triplet superconductivity driven by quantum-critical fluctuations.
\end{abstract}
\maketitle

%
%
	

	Superconductivity
	(SC) in	moir\'e graphene, occurring in proximity to other correlated electronic orders\cite{Bistritzer12233,cao2018SC,lu2019superconductors,andrei2020graphene,saito2020independent,polshyn2019large,Oh2021,Cao_2021,jaoui2022quantum}, attracts intense interest as a possible instance of an unconventional (Coulomb-repulsion-driven) pairing mechanism. However, so far no consistent picture has emerged. In some cases SC appears to be associated with correlated orders \cite{Cao_2021,jaoui2022quantum}, suggesting exotic pairing scenarios\cite{you2019superconductivity, Kozii2019,Chichinadze2020, khalaf2021charged}, whereas
	other experiments report on SC that can be isolated (and thus decoupled) from
 other
 ordered phases
	\cite{saito2019decoupling,saito2020independent}.
	Presently, there is no clear verdict
	on the pairing mechanism, as some studies point to the irrelevance of electron interactions
	for the pairing \cite{stepanov2020untying, polshyn2019large}, supporting the conventional phonon
mechanism
\cite{Wu2018, Lian2019,Wu2019}, whereas others suggest these interactions
	do
	matter\cite{liu2021tuning}. System complexity, such as 
	the peculiar form of moir\'e flatband electron wavefunction\cite{Bistritzer12233,Tarnopolsky2019} and multiple kinds of moir\'e-related 
	disorder (e.g. twist-angle disorder, strain, buckling, and so on), make this debate difficult to settle.
	
Helpfully, however, recent work unveiled
two non-moir\'e systems that host superconducting orders intertwined with correlated electronic orders --- the field-biased rhombohedral trilayer graphene\cite{zhou2021RTG,zhou2021superconductivity} (RTG) and Bernal bilayer graphene\cite{zhou2021BLG,de2021cascade} (BBG). These systems present a distinct advantage for studying strongly-correlated physics owing to the simplicity of their bandstructure, tunable by an external transverse electric field, and exceptional cleanness
due to the absence of strain. Here we focus on BBG, where a particularly interesting superconducting order
 has been observed
 \cite{zhou2021BLG}.
BBG is a system with a simple band structure 
 consisting of two bands \cite{McCann_2013,McCann2006Landau}, whose wavefunctions are predominantly constructed from the orbitals in one of the two layers:
 one
   from
     the
     A sublattice in 
      the 
      upper layer, the other 
       from the B sublattice in
         the 
         bottom layer. In the absence of transverse field, the two bands have quadratic dispersion and touch at the high-symmetry 
           points
           $K$ and $K'$. In unbiased BBG, theory anticipates various correlated ordered states 
           \cite{Nandkishore2010dynamical,Nandkishore2010quantum,Vafek2010,Jung2011lattice,macdonald2012pseudospin,Zhang2012Distinguishing,Cvetkovic2012,Throckmorton2014,Min2008,Nilsson2006}, some of which have been identified in experiments\cite{martin2010local, weitz2010broken, Mayorov2011, Veligura2012, velasco2012transport,Bao2012,Freitag2012}.

Recent renewal of interest in BBG is triggered by the experimental realization of a flatband regime in this system, where new exotic orders are found\cite{zhou2021BLG,de2021cascade,Seiler2021}. 
This new regime is accessible through applying a transverse electric field to
   open a band gap at charge neutrality 
   \cite{McCann2006Landau,McCann2006Asymmetry}.
Under an applied field band dispersion changes from quadratic to quartic, which flattens out as the field grows. 
This leads to properties completely distinct from those of unbiased bilayer graphene \cite{Nandkishore2010dynamical,Nandkishore2010quantum,Vafek2010,Jung2011lattice,macdonald2012pseudospin,Zhang2012Distinguishing,Cvetkovic2012,Throckmorton2014,Min2008,Nilsson2006}. Specifically,
as seen in experiments\cite{zhou2021BLG,de2021cascade,Seiler2021},
at a low carrier density
 BBG exhibits an isospin instability,
  where electrons populate only one or two isospin 
   sub-bands 
   [phases PIP$_1$ and PIP$_2$ in 
    Fig.\ref{fig:PD} a), 
     where isospin refers to spin and $K/K'$ 
     valley degrees of
freedom].
 This behavior leads to
a cascade of phase transitions between states with different polarization, resembling those 
 in moir\'{e} graphene\cite{saito2021isospin,zondiner2020cascade,rozen2021entropic,choi2021correlation,pierce2021unconventional} and in rhombohedral trilayer graphene\cite{zhou2021RTG}.
 On top of this cascade of isospin orders, other symmetry-breaking orders are predicted theoretically, e.g. the momentum space polarization wherein all carriers are shifted into one, two or three pockets at the band minima produced by the trigonal warping effects\cite{Jung2015,dong2021}.

Further, both BBG and RTG host superconductivity
\cite{zhou2021RTG,zhou2021BLG}. In both BBG and RTG the superconducting phases occur at phase boundaries between different isospin-ordered states, forming narrow sleeves extending along the phase boundaries.
There are interesting differences between these SC phases. In RTG, there are two SC phases in the hole-doped regime\cite{zhou2021BLG}, which show a conventional suppression under an applied magnetic field.
To the contrary, in BBG, there is only one SC phase (the cyan area in Fig.\ref{fig:PD} c) which arises 
at the phase boundary between isospin-ordered and isospin-disordered phases 
in the presence of an in-plane magnetic field. At zero magnetic field, superconductivity is suppressed, giving way to a 
correlated state with an insulator-like temperature dependence of resistivity. This state, of a yet unknown origin, 
is seen as a red dome in 
 panels b) and c) in 
 Fig.\ref{fig:PD}.
In the literature, several candidate mechanisms for SC in RTG have been proposed \cite{Ghazaryan2021,chatterjee2021inter,You2022}, however the origin of SC and the adjacent correlated phase in BBG have not yet been understood.

	\section{Exotic superconductivity in Bernal bilayer graphene }

\begin{figure}[t]
	\centering
	\includegraphics[width=0.48\textwidth]{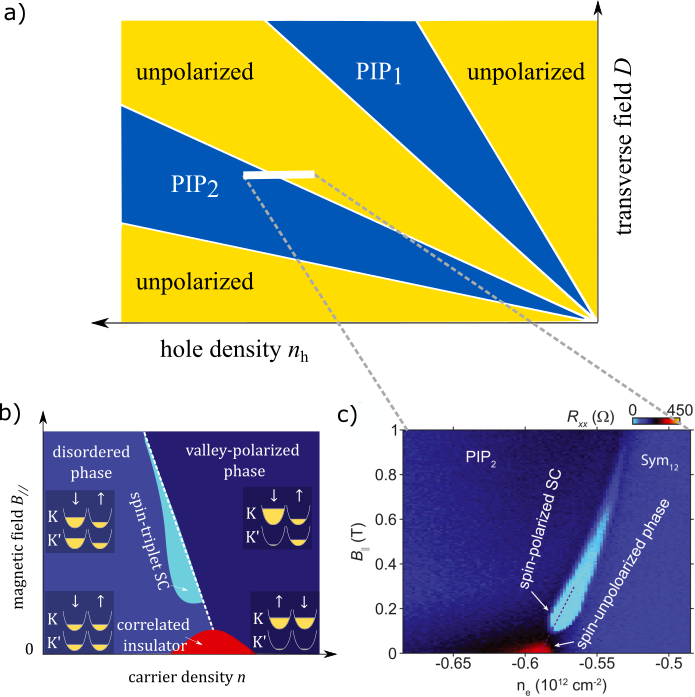}
	\caption{a) Schematic of the experimental isospin-order phase diagram 
	[adapted from Ref.\cite{zhou2021BLG}]. In phases PIP$_1$ and PIP$_2$ only one or two isospin-polarized 
 sub-bands 
  are populated.
Here, we focus on the isospin polarization in each phase, ignoring other differences between unpolarized and partially polarized states that are irrelevant for our discussion. 
b) Predicted phase diagram for superconductivity governed by critical mode at the phase boundary between isospin-valley-polarized and unpolarized phases. Superconducting order is spin-triplet and is induced by a finite magnetic field that creates spin imbalance as shown in the insets. The pairing interaction in the triplet channel is a repulsion at $B=0$, turning into an effective attraction at a finite $B$. c) The experimental phase diagram for superconductivity in bilayer graphene\cite{zhou2021BLG} strongly resembles the phase diagram predicted in b) [see text].}
	\vspace{-5mm}
	\label{fig:PD}
\end{figure}
	
Perhaps the most mysterious and intriguing 
phenomenon observed in BBG systems
is the magnetic-field-induced SC, which will be the subject of this study. 
  Ref.\cite{zhou2021BLG} reports on
	a unique dependence of SC on parallel magnetic field, suggesting a pairing mechanism entirely different from those studied in moir\'e \cite{you2019superconductivity, Kozii2019,Chichinadze2020, khalaf2021charged} and
	RTG systems \cite{zhou2021RTG,Ghazaryan2021,Chou2022}.
	Specifically, it is found that SC occurs only when $B_\parallel$ exceeds a finite threshold
 (see Fig.\ref{fig:PD}c).
  This is quite unlike textbook SC,
   which is suppressed by a $B$ field.  Since $B_\parallel$ only couples to spin when applied in-plane, the $B$-induced SC indicates that spin imbalance is essential for pairing.
	Moreover,
		SC is found to persist in a high field,
		surviving well above the Pauli limit.
  The resilience of SC in a $B$ field unambiguously points to a spin-triplet pairing, and thus an unconventional pairing mechanism.
	
Some clues for pairing mechanism are revealed by
 several features,
  shared by BBG and RTG. In both systems, SC tracks the boundary between isospin
  ordered
   and disordered  phases
 (PIP$_2$ and Sym$_{12}$  phases in the notations of \cite{zhou2021BLG}). This boundary
 remains sharp in the presence of a $B$ field \cite{zhou2021RTG,zhou2021BLG}.
	This suggests
	pairing
 mediated by a critical
	isospin mode --  a scenario,
in which both
	superconductivity
	and
	the Stoner
	instability
	responsible for isospin
	order
	arise from electron-electron repulsion.
	A
 pairing of this type would place BBG into the
  class of systems  with quantum-critical SC \cite{Klein2020,Oganesyan2001,Lederer2015,Tremblay2013,Chubukov_2020a,berg_4,sslee_2018,efetov2013}.

Yet, this scenario, encounters a crucial obstacle when applied to BBG system. In previously studied instances, soft quantum-critical 
modes generate an effective e-e attraction either by scattering Cooper pairs between different Fermi surfaces
 or different hotspots on a single Fermi surface\cite{Mazin2008unconventional,Mazin2009pairing}
  by exchange-type interactions \cite{Scalapino1966}.
   This does not work for BBG because
 Cooper pairs in graphene are formed by electrons in valleys $K$ and $K'$,
   and the size of a Fermi surface in each valley
  is much smaller than the reciprocal lattice vector.
This excludes the usual pair hopping scenarios and makes the exchange-type processes negligible
(see Ref.\cite{SM} for more detail).
 As the consequence,
 the soft-mode-mediated interaction, while being
 strong, is repulsive in graphene systems. Therefore, understanding this unusual superconductivity 
 requires a new mechanism that converts a strong repulsion,
  mediated by the soft modes,
   into an attraction.
	
\begin{figure}[t]
	\centering
	\includegraphics[width=0.49\textwidth]{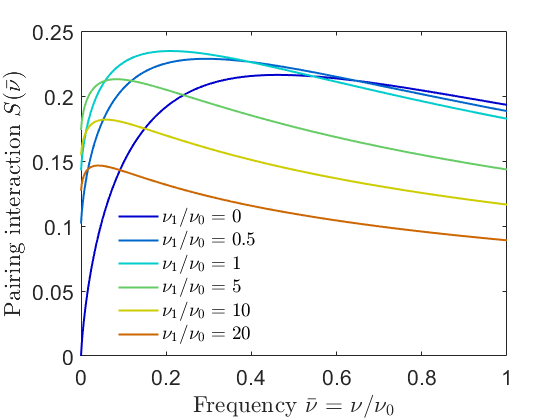}
	\caption{
		Frequency-dependent pairing interaction $S({\bar
			\nu})$, Eq. (\ref{eq:gap equation_1}), describing the universal contribution of a quantum-critical mode.
		The nonmonotonic frequency dependence, which is
		a
		generic property of
		$S({\bar \nu})$
		for all
		values of the stiffness $K$, allows for a repulsive interaction to generate an attractive effective pairing interaction. The value at $\nu=0$ is finite at $K>0$ and zero at $K=0$. In the latter case superconducting $T_c$
		is non-zero for
		any $\delta_{\downarrow}$, the distance to critical point for spin-down fermions.}
	\label{fig:S}
\end{figure}

Here we propose
  such
   mechanism for
   BBG. This mechanism
    ties together
all ingredients mentioned above:
a quantum-critical mode, a
 repulsive
 Coulomb coupling, a broken spin degeneracy, and pairing in spin-triplet channel.
 We find that the pairing interaction
 acquires
 new properties at a finite
  spin
  imbalance,
   which
   lead to a
    SC instability.
First, for spins opposite to $B$, the pairing interaction
at Stoner transition,
while remaining repulsive at all bosonic frequencies
 $\nu$,
  acquires a non-monotonic frequency dependence.
    Namely, it  drops sharply at small
    $\nu$
    and passes through a maximum at intermediate $\nu$
 (see Fig.\ref{fig:S}). This behavior is ``universal'' in the sense that the existence of such non-monotonicity is independent of the details in the bandstructure.
 This non-monotonicity enables pairing with a superconducting gap $\Delta (\omega)$ changing sign between small and large $\omega$ \cite{Morel1962,McMillan1968,Bogoljubov1958,Coleman2015,prokofiev}.
We show that in our case this mechanism leads to
spin-triplet, valley-singlet $s-$wave pairing.
Second, for spins along $B$, the  same
   fully dressed
   pairing interaction, mediated by a quantum-critical mode,  acquires field-induced dependence on soft-mode momenta. This generates finite interaction components in non-$s-$wave channels.
We show that in our case this  leads to pairing in spin- and valley-triplet,  spatially-odd $p$-wave channel.
 This is essentially a field-induced Kohn-Luttinger (KL) mechanism \cite{Kohn1965} in 2D.

We emphasize that the attraction for both spin components comes from the same effective interaction mediated by soft isospin excitations, i.e., the two attractive channels are intertwined.
Either  pairing mechanism
 explains all the qualitative features in the experimentally measured phase diagram, 
 Fig.\ref{fig:PD} c). First,
superconductivity
peaks at the phase boundary.
Second, superconductivity has a threshold in $B$ field as
one needs a finite $B$ to overcome initial repulsion.
At $B$ below the threshold we predict a
strong repulsive interaction. This interaction is expected
to produce a correlated insulator state, in line with experiment [Fig.\ref{fig:PD} c)].

One more aspect of the observed phase diagram that supports this scenario is the dependence of the onset of the PIP$_2$ order on the in-plane $B$ field.
In experiments,
the $B$ field merely pushes the
onset of the order
to lower carrier densities, producing a phase boundary with a constant slope marked by dashed line in Fig.\ref{fig:PD} c).
As we will see, this behavior follows directly from
our model (see \eqref{eq:phase_boundary_slope}
below
and accompanying discussion).
We note that in Fig.\ref{fig:PD} c), the $x$ axis is drawn from high carrier density to low carrier density, and must be reversed when compared to panel b).
The experimental and theoretical slopes of the phase boundary match
both in sign and in value, lending further support to theory.

\section{The pairing interaction due to quantum-critical modes}
We now
proceed with the analysis. We model the
interacting electrons in BLG using the conventional
Hubbard
Hamiltonian:
\be
H = \sum_{i} \epsilon_{i }(\vec p) \psi_{i \vec p}^{\dagger} \psi_{i \vec p} + \sum_{ii'} \frac{V_{0}}{2} \psi_{i \vec p+\vec q}^\dagger \psi_{i' \vec p'- \vec q}^\dagger \psi_{i' \vec p'}\psi_{i \vec p}
,
\ee
where $i,i'= K\uparrow, K\downarrow, K' \uparrow, K'\downarrow$ are isospin indices,
$V_{0}$ is a short-range repulsion. The band structure $\epsilon(\vec p)$ details will be inessential in either mechanism we are going to describe in next two sections.
 Specifically, as we will see later,  the role of $\epsilon(\vec p)$ in our first mechanism can be described by one parameter, and only affects the result quantitatively, whereas in the second mechanism, a realistic $\epsilon(\vec p)$ is only a small twist on the case of parabolic dispersion which we will use as a minimal model. This small twist leads to a finite field threshold, but does not invalidate the scenario.
Since they are inessential, we will 
keep it general for now and specify them later when needed.

\begin{figure}[t]
	\centering
	\includegraphics[width=0.48\textwidth]{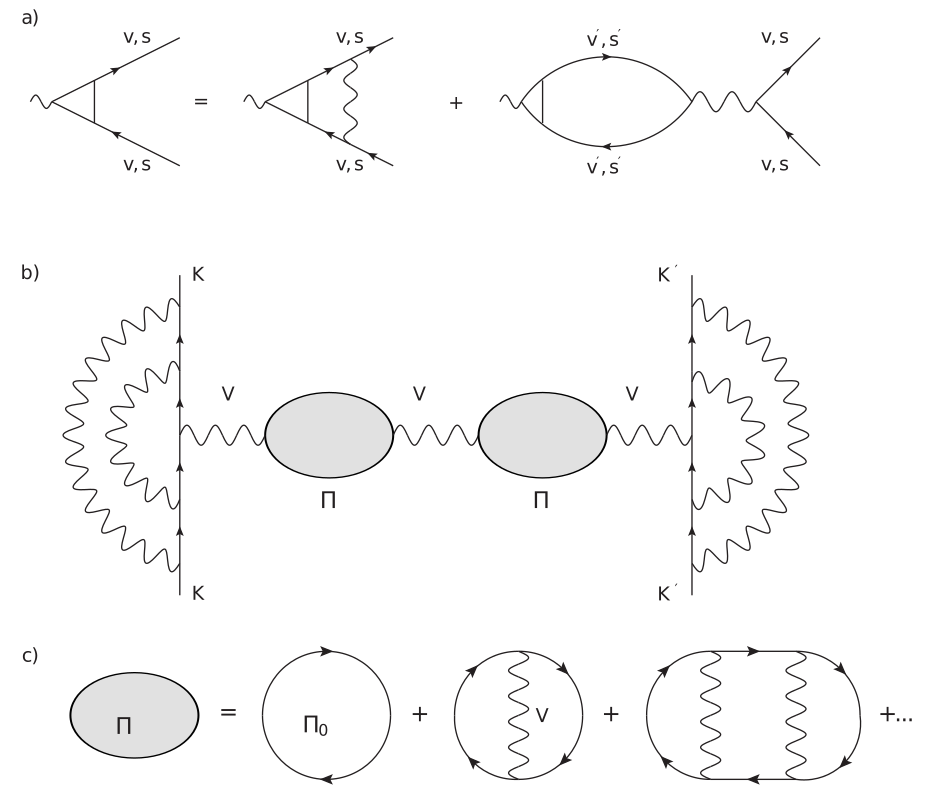}
	\caption{ a) The RPA diagrams describing the valley-polarization instability. Here $v,s$ are valley and spin indices. b) Diagrams describing the effective pairing interaction between electrons in valleys $K$ and $K'$ mediated by quantum-critical modes.
		These processes give a divergent enhancement to forward scattering near the valley-polarization instability (see text).
		The
		analytical expression for the effective interaction
		is given in \eqref{eq:H_eff1}. c) A diagrammatic description of the renormalized susceptibility.
	}\label{fig:leading diagram}
\end{figure}

A conventional RPA-type analysis of interaction-induced particle-hole instabilities
  shows
that  in a field the system develops an intra-valley spin or charge $q=0$ order, which changes sign between the two valleys (this is what we termed as an isospin valley-polarized order). The condition for the instability is $1+ V_0 \Pi_{0,s} (0,0) =0$, where $s =\uparrow$ or $\downarrow$ with respect to $B$, and  $\Pi_{0,s} (\nu, q)$ is a bare polarization bubble (see Fig.\ref{fig:leading diagram}).
 Here $\nu$ and $q$ denote
Matsubara frequency and momentum transfer, and
  $\Pi_{0,s} (0,0)$, as defined, is negative.  Below we assume, following the experiments, that $|\Pi_{0,\uparrow} (0,0)| >
|\Pi_{0,\downarrow} (0,0)|$, hence the first instability at a non-zero $B$ is for fermions with spin parallel to $B$.

The pairing interaction
 involves fermions with  momenta ${\bf k}$ and $-{\bf k}$,  which in our case belong to different valleys
 $K$ and $K'$.  At the lowest order, the pairing interaction is just $V_0$, however near the onset of an isospin order
dressing of the pairing interaction by particle-hole bubbles is essential.  We argue \cite{SM}
 that the relevant diagrams are the ones
shown in Fig.\ref{fig:leading diagram}, where each arrow represents the electron's Green's function $G_{
	s}(\omega, \vec p) = 1/\lp i\omega -\epsilon_{
	s}(\vec p) \rp$.
The resulting effective interaction can be written as
\bea
&& \Gamma_{s
	s}(\nu, q) =
\gamma^2_{
	s}(\nu,q)V(\nu,q),
~~
\gamma_{
	s}(\nu, q) = \frac{1}{1+V_0\Pi_{0,
		s}(\nu, q)},
\nonumber \\
&& \quad V(\nu,q) = \frac{V_0}{1- 2
	V_0 \sum_{s'} \Pi_{0,s'} (\nu,q) \gamma_{s'} (\nu, q)}.
\label{eq:H_eff1}
\eea
Here and below we approach SC from a non-polarized state and assume
valley symmetry.
Near the onset of valley polarization, $V_0 \Pi_{0,s} (\nu, q=0) \approx -1$, and
at small  momentum transfer $q\ll k_F$
and small $\nu < v_F q$
the vertex-correction
factors $\gamma_{s}(\nu,\vec q)$ take
the familiar form:
\be
\gamma_{s}
(\nu,\vec q) \sim \frac{1}{\frac{|
		\nu
		|}{v_F q} + Kq^2+ \delta_s}
		,
\label{eq:gamma}
\ee
where $\delta_s$ is a distance to valley-polarization phase boundary for fermions with a given spin projection.
The stiffness $K$ and Fermi velocity $v_F$,
 taken here
 to be spin-independent,
are determined by band dispersion $\epsilon(\vec p)$.
At zero $B$, $\delta_\uparrow=\delta_\downarrow$, the effective interaction is
\be
\Gamma_{ss}(
 \nu,q) =
\frac{V_0 \gamma(\nu,\vec q)^2}{1-4 V_0 \Pi_0(\nu, q)\gamma(\nu,\vec q)} \approx  \frac{1}{4}\frac{V_0}{\frac{|
		\nu|}{v_F q} + Kq^2+ \delta}.\nonumber
\ee
The sign of this interaction is a repulsion, in distinction to
 that found for pairing mediated by a critical
$q=0$ mode for
QHE composite fermions \cite{Bonesteel1996} and for a
nematic QCP\cite{Klein2020,Oganesyan2001,Lederer2015}. In these systems electrons
with ${\bf k}$ and ${\bf -k}$ live on the same Fermi surface and
interact through an exchange
processes. In our case, such a process is forbidden as it requires a fermion to scatter from one valley to the other.
However, a nonzero $B$ field lifts spin degeneracy and  generates
new frequency and momentum scales below which the interaction $\Gamma_{ss} (\nu, q)$ gives rise to pairing,
  as we now demonstrate.

\begin{figure}[t]
	\centering
	\includegraphics[width=0.5\textwidth]{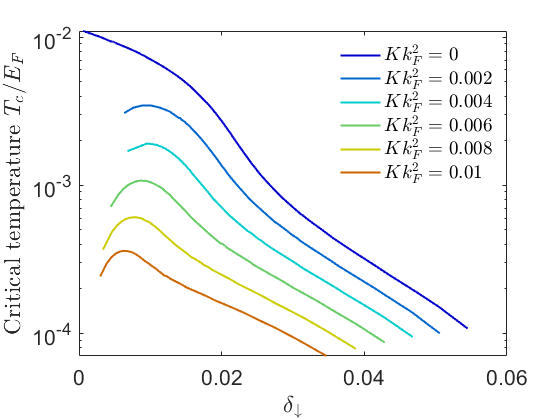}
	\caption{
		Critical temperature  $T_c$ 
		vs. the distance to criticality $\delta_{\downarrow}$ for down spins, a parameter controlled by magnetic field. 		For $K=0$, $T_c $ is nonvanishing for all $\delta$ (see text). For $K\neq 0$, to the contrary, each curve starts at a finite threshold value $\delta_{\downarrow}>0$. As $K$ grows, the threshold value first grows and then decreases, reflecting the behavior of $S(\nu)$ at small $\nu$ shown in Fig.\ref{fig:S}.
	}\label{fig:Tc}
\end{figure}

\section{ Valley-singlet $s$-wave pairing}
A possibility of an $s-$wave pairing from nominally repulsive (positive) interaction  has been discussed several times in the literature both some time ago \cite{Morel1962,McMillan1968,Bogoljubov1958,Coleman2015} and recently \cite{ruhman,prokofiev,pimenov}.  The idea is that if the interaction either has a non-monotonic frequency dependence or is reduced at small frequencies, there may emerge an effective attraction for a gap function that changes sign between small and large frequencies, in analogy with how a nominally repulsive interaction may become attractive in a
non-$s-$wave spatial channel where a gap changes sign between different regions on the Fermi surface.
In our case, at zero field the average interaction
 $\overline\Gamma_{ss} (\nu)$, which is defined as
  $\Gamma_{ss} (\nu,\vec q)$  averaged over momentum transfers $\vec q$ on the Fermi surface, is a monotonically decreasing function of frequency $\nu$, and a solution with  sign-changing $\Delta (\omega)$ is impossible.
At a finite $B$,
 realistic
bandstructure calculations  in Ref.\cite{zhou2021BLG} show that the valley order sets first for majority spin (spins along $B$), and near the onset  of the order,  $\delta_{\uparrow} \ll \delta_{\downarrow}$
 and $\gamma_{\uparrow}\gg \gamma_{\downarrow}$.
In this situation,
 the average interaction for spin-up electrons
$\overline\Gamma_{\uparrow \uparrow} (\nu)$ still monotonically decreases with $\nu$, but the average interaction for spin-down electrons
$\overline\Gamma_{\downarrow \downarrow} (\nu)$ becomes non-monotonic.  This happens because $\Gamma_{\downarrow \downarrow} (\nu, q)$ has a linear rather than inverse linear dependence on $1+ V_0 \Pi_{\uparrow} (\nu, q)$.
In explicit form, the averaged interaction $\overline{\Gamma}_{\downarrow \downarrow} (\nu)$ is
\be
\overline{\Gamma}_{\downarrow \downarrow} (\nu) = V_0 \int_0^{k_F}  \frac{dq}{2\pi} \frac{\frac{|\nu|}{v_F|q|} +Kq^2}{\lp \frac{|\nu|}{v_F|q|} +Kq^2 + \delta_{\downarrow} \rp^2}
\label{eq:n1}
\ee
The result can be cast into the scaling form $\overline\Gamma_{\downarrow \downarrow} (\nu) = V_0 (k_F/2\pi \delta_{\downarrow}) S\left(\nu/\nu_0,  \nu_1/\nu_0\right)$, where $\nu_0 = 2 E_F \delta_{\downarrow}$, $\nu_1 = 2 E_F (K k^2_F)$, and $E_F = v_F k_F/2$.   To see the consequence, consider first the case $\nu_1 \ll \nu_0$. The
 function
$S(x,0) = x (\log{\frac{1+x}{x}} -\frac{1}{1+x})$.
 is  manifestly non-monotonic: it is linear in $x$ at small $x$, passes through a maximum at $x \sim 0.5$, and drops at higher $x$. We emphasize that the non-monotonic behavior is fully induced by $B$, which splits $\delta_{\uparrow}$ and $\delta_{\downarrow}$.  The non-monotonic dependence holds if we increase the ratio $\nu_1/\nu_0$
   as we show in  Fig.\ref{fig:S}.

The gap equation for the pairing of spin-down fermions, mediated by $\overline{\Gamma}_{\downarrow \downarrow} (\nu)$, is
\be
\Delta(\omega) = - \frac{T_c}{2 v_F}   \sum_{\omega'=\pi T_c (2n+1)} \frac{\Delta(\omega') \overline{\Gamma}_{\downarrow \downarrow} (\omega - \omega')} {|\omega'|},\label{eq:gap equation}
\ee
The overall minus sign reflects that the interaction is repulsive.
The gap equation takes a universal form when expressed in terms dimensionless ${\bar T}_c = T_c/\nu_0$ and ${\bar \omega} = \omega/\nu_0$:
\be
\Delta({\bar \omega}) = - \lambda  \pi {\bar T}_c   \sum_{\bar{\omega}'=\pi {\bar T}_c (2n+1)} \frac{\Delta({\bar \omega}')} {|\bar{\omega}'|}
S\left({\bar \omega} -{\bar \omega}',0\right)
\label{eq:gap equation_1}
\ee
where $\lambda = k_F V_0/(4\pi^2 v_F \delta_{\downarrow})$.
Because $S \left({\bar \omega} -{\bar \omega}',0\right)$ is strongly peaked at $|{\bar \omega} -{\bar \omega}'| ={\bar \nu}_* \approx 0.5$, one can change the overall sign in (\ref{eq:gap equation_1}) by searching for gap functions which change sign under
${\bar \omega} \to {\bar \omega} + {\bar \nu}_*$.
 At small $\lambda$, analytical consideration yields $T_c \propto \omega_0 e^{-1/\lambda^2}$ (Ref.\cite{prokofiev}). At  $\lambda \leq 1$, $T_c \sim \omega_0$, but with a numerically small prefactor.
 At larger $\lambda$, the prefactor increases and at $\lambda \gg 1$ (i.e., at small $\delta_{\downarrow}$,
  $T_c \sim \lambda \nu_0 \sim E_F$. For full consideration, at large $\lambda$ one also
  has to include  fermionic self-energy.
  In supplement, we estimate the effect of self-energy generated by the pairing interaction, and find that the resulting suppression of $T_c$ is acceptable. However, we emphasize that the self-energy comes from intravalley interaction, which is different from the intervalley pairing interaction. They only coincide with each other in mean-field theory. In reality, the self-energy does not have to diverge when pairing interaction diverge. Therefore, it is safe to ignore the role of self-energy.
   We show $T_c$, obtained by numerical solution of Eq. (\ref{eq:gap equation_1}), in
 Fig.\ref{fig:Tc}.
   We set
    $k_F V_0/(2\pi v_F) = 1$,
      as required for a Stoner  instability, and set $E_F = 10 \rm{meV}$.  We see that at $\nu_1=0$,
       $T_c$ monotonically increases with decreasing $\delta_{\downarrow}$ and at small $\delta_{\downarrow}$ saturates at
        roughly $1\rm{K}$.

  For a more realistic case of  $\nu_0 \sim \nu_1$, the momentum-averaged  $\overline\Gamma_{\downarrow \downarrow} (\nu)$ tends to a finite value at $\nu =0$, leading to a smaller $T_c$
     and also setting a threshold on $\delta_{\downarrow}$ as an $s-$wave pairing by a frequency-dependent repulsion is a threshold phenomenon \cite{ruhman,prokofiev,pimenov}, and at a small $\delta_{\downarrow}$ the non-monotonicity of ${\overline \Gamma}_{\downarrow\downarrow} (\nu)$ is too weak to give rise to a pairing when the self-energy is included.  At larger $\delta_{\downarrow}$  $T_c$ also  drops because the coupling $\lambda$ gets smaller.
    This gives rise to a dome-like dependence of $T_c$ on
     $\delta_\downarrow$ at a given
 $Kk_F^2$.
  For $Kk_F^2 = 10^{-2}$
   we obtained at $ T_c \sim 35\rm{mK}$
   at optimal $\delta_\downarrow  = 7\times 10^{-3}$.
This value is in line with experimental $T_c$.

\section{Valley-triplet $p$-wave pairing.} We now argue that at a finite $B$, the effective interaction, mediated by soft isospin fluctuations,  also gives rise to  an attraction in another spin-triplet channel, this time valley triplet and spatially odd. The mechanism here is field-induced KL effect in 2D -- the development of attractive spatial  component(s) due to screening of a purely repulsive bare pairing interaction by particle-hole polarization bubbles.
This pairing comes from momentum transfers $q \sim 2k_F$, when there is no good theoretical reason to  restrict with only diagrams in Fig. (\ref{fig:leading diagram}).  We assume without proof that Eq. (\ref{eq:H_eff1}) is still valid, at least by order of magnitude,  when $q \sim 2k_F$.

To understand field-induced KL effect, assume momentarily parabolic $\epsilon_i (p)$  near $K$ and $K'$ and consider static interaction $\Gamma_{ss} (0, q)$.
The free-fermion  polarization in 2D is $\Pi_s (0,q) = -(m/2\pi)$ for $q < 2k_{F,s}$ and $-(m/2\pi) \left(1 -
\sqrt{4 k^2_{F,s}/q^2}\right)$ for $q > k_{F,s}$.  Relevant $q$ for SC are below $2k_{F,s}$ for a given spin projection. At $B =0$,  $k_{F,s} = k_F$ is the same for up- and down-spins. In this situation, $\Pi_s (0,q) = -m/(2\pi)$ for relevant $q$, and the effective interaction $\Gamma_{ss} (0,q)$  has only an $s$-wave repulsive component, like the bare $V_0$.  This is commonly known as the absence of KL effect in 2D for a parabolic dispersion \cite{baranov1992}.
 The situation changes at a finite $B$ as now the effective interaction for  fermions with spin-up partly comes from
fermions with spin-down and vice versa. Because the Fermi momentum $k_{F,\uparrow}$ is larger than $k_{F,\downarrow}$, there is a range $2k_{F,\downarrow} < q < 2k_{F,\uparrow}$, where the interaction
$\Gamma_{\uparrow\uparrow} (0, q)$ for spin-up fermions at momentum transfer on their Fermi surface acquires a momentum-dependence via the momentum dependence of $\Pi_{\downarrow} (0, q)$.  There is no such effect for
$\Gamma_{\downarrow\downarrow} (0, q)$ at $q < 2k_{F,\downarrow}$.

Once $\Gamma_{\uparrow\uparrow} (0, q)$ becomes momentum-dependent, one can search for
spatially-odd
 solutions
$\Delta (\theta)$, subject to $\int d \theta \Delta (\theta) =0$ and $\Delta (\theta + \pi)=-\Delta (\theta)$,  where $\theta$ is an angle along the Fermi surface counted from, e.g., $k_x$ direction.  These gap functions are necessary valley-triplets.  The analysis of the pairing instability is rather standard and we just present the result.
 We find that the $q-$dependence of the interaction gives rise to an attraction for spatially-odd $\Delta (\theta)$. At at a small field  the gap equation is approximately local in $\theta$, and
yields
$T_c \sim E_F e^{-1/\lambda_{KL}}$, where
\be
\lambda_{KL} = \frac{m V_0}{8\pi^2 \delta^2} \frac{\mu_B B}{E_F}
\label{eq:KL}
\ee
We see that the field-induced
 $\lambda_{KL}$
  is positive. At small $B$,  $T_c$ increases exponentially with
the field. At a larger $B$, the prefactor gets smaller as the number of down-spin fermions decreases.  As a result, $T_c$ has a dome-like shape as a function of $B$.
We also note that Eq. (\ref{eq:KL}) is valid when
$\lambda_{KL} <1$. At larger coupling, the coupling gets renormalized by fermionic self-energy and eventually saturates.
 The analysis can be straightforwardly extended to the physically relevant case  $\delta_{\uparrow} \ll \delta_{\downarrow}$, which one can model by non-equal DOS for up and down spins.  We found that Eq. (\ref{eq:KL}) holds, but $\delta^2$
 in (\ref{eq:KL})
  has to be replaced by $\delta^2_{\downarrow}$.  One can also   move away from parabolic dispersion and include the $q$-dependence of $\Pi (0, q)$ at $q <2k_F$.  Similarly to valley-singlet case,
this will (i) decrease $T_c$ and (ii) set a finite threshold on a field as field-induced attraction has to compete with
 a
   repulsive bare interaction in  valley-triplet channel.  As a result, $T_c$ as a function of $B$ displays a dome-like behavior above a finite threshold, much like for valley-singlet $s-$wave pairing.

\section{Relation to experiments}
We now discuss several items related to experiments.
First, in our analysis, particularly of valley-singlet SC, we assumed that isospin order sets up first for spin-up fermions.
To see that our understanding of the phase transition is correct, below we calculate the slope of the phase boundary, and comparing it with experiment. According to our model, instability happens only in majority spin. Therefore, at the phase transition, the density of carriers in the majority spin is a fixed value, while the density of carrier in minority spin depends on $B$ field. Specifically, we expect that the phase transition shifts towards lower total carrier density under increasing $B$. The shift of carrier density is linear in magnetic field:
\be\label{eq:phase_boundary_slope}
\frac{dn_*}{dB} = 2\mu_B \nu_0,
\ee
where $n_*$ is the total carrier density at the phase boundary, $\mu_B$ is the Bohr magneton, $\nu_0$ is the density of states per isospin, the factor of $2$ arises from $K$/$K'$ valley degeneracy. Plugging in the value of the density of states obtained numerically in Ref.\cite{zhou2021BLG}, we find the slope is $\frac{dn_*}{dB} = 5\times 10^{-4} \rm{nm}^{-2} \rm{meV}^{-1}$, which matches the slope extracted from Fig.\ref{fig:PD}c).

 Second, the
  two pairing scenarios that we discussed  yield  dome-shaped $T_c (B)$ with a threshold on $B$,
but differ in which spin components pair:
valley-singlet pairing involves  spins opposite to the field, while
valley-triplet pairing involves spins along the field.
One way to  test which spin components are  involved in SC
is to measure the DC voltage drop when injecting a spin-polarized current into the system. If we inject electrons from a ferromagnetic material which is polarized by the same in-plane magnetic field as in the BLG, then
our theory predicts that for valley-singlet pairing this spin-polarized current
should
give
a finite DC voltage drop even
$T_c$ because the SC only occurs in the Fermi sea of the opposite spin polarization, while for valley-triplet pairing a DC voltage drop should disappear below $T_c$.

Third, valley-singlet pairing arises from small-momentum part scattering,
and  should be sensitive to screening. When a metallic gate is brought closer to the sample, the Coulomb interaction will be suppressed. As a result, the width (in terms of density) of the SC phase should narrow when a proximal metal gate is applied, because the same value of $T_c$ can be achieved only by getting closer to the phase boundary.    For valley-triplet pairing, this effect is smaller as the pairing is not confined to small-momentum scattering.

Finally, can the
quantum-critical mode contribute to resistivity through carrier scattering by thermal fluctuations? This may seem plausible at a first glance, yet
in our scenario a strong effective interaction, mediated by a soft boson, holds for small momentum transfers $q\ll k_F$
due to the proximity to
the $q=0$ isospin order.
As a result, despite thermal fluctuations being strong, forward scattering does not produce a relaxation of current and thus does not contribute to resistivity \cite{Maslov2011}.
This is in line with the experiment where $T$ dependence of resistivity shows no signature of critical fluctuations near the phase transition\cite{zhou2021BLG}.

We therefore conclude that all the unique aspects of the observed superconductivity are successfully explained by the attraction-from-repulsion-based pairing scenario. Furthermore, this mechanism is `natural' as it arises from the strong electron-electron interactions that drive
the adjacent isospin-polarized electron orders.
 As such, it constitutes a unique verifiable instance of exotic pairing. Supported by experiments, it sheds light on the origin of spin-triplet superconductivity in BBG and is applicable to a variety of other systems of interest.

\bibliography{ref}

\begin{thebibliography}{79}
\expandafter\ifx\csname natexlab\endcsname\relax\def\natexlab#1{#1}\fi
\expandafter\ifx\csname bibnamefont\endcsname\relax
  \def\bibnamefont#1{#1}\fi
\expandafter\ifx\csname bibfnamefont\endcsname\relax
  \def\bibfnamefont#1{#1}\fi
\expandafter\ifx\csname citenamefont\endcsname\relax
  \def\citenamefont#1{#1}\fi
\expandafter\ifx\csname url\endcsname\relax
  \def\url#1{\texttt{#1}}\fi
\expandafter\ifx\csname urlprefix\endcsname\relax\def\urlprefix{URL }\fi
\providecommand{\bibinfo}[2]{#2}
\providecommand{\eprint}[2][]{\url{#2}}

\bibitem[{\citenamefont{Bistritzer and MacDonald}(2011)}]{Bistritzer12233}
\bibinfo{author}{\bibfnamefont{R.}~\bibnamefont{Bistritzer}} \bibnamefont{and}
  \bibinfo{author}{\bibfnamefont{A.~H.} \bibnamefont{MacDonald}},
  \bibinfo{journal}{Proceedings of the National Academy of Sciences}
  \textbf{\bibinfo{volume}{108}}, \bibinfo{pages}{12233}
  (\bibinfo{year}{2011}), ISSN \bibinfo{issn}{0027-8424},
  \eprint{https://www.pnas.org/content/108/30/12233.full.pdf},
  \urlprefix\url{https://www.pnas.org/content/108/30/12233}.

\bibitem[{\citenamefont{Cao et~al.}(2018)\citenamefont{Cao, Fatemi, Fang,
  Watanabe, Taniguchi, Kaxiras, and Jarillo-Herrero}}]{cao2018SC}
\bibinfo{author}{\bibfnamefont{Y.}~\bibnamefont{Cao}},
  \bibinfo{author}{\bibfnamefont{V.}~\bibnamefont{Fatemi}},
  \bibinfo{author}{\bibfnamefont{S.}~\bibnamefont{Fang}},
  \bibinfo{author}{\bibfnamefont{K.}~\bibnamefont{Watanabe}},
  \bibinfo{author}{\bibfnamefont{T.}~\bibnamefont{Taniguchi}},
  \bibinfo{author}{\bibfnamefont{E.}~\bibnamefont{Kaxiras}}, \bibnamefont{and}
  \bibinfo{author}{\bibfnamefont{P.}~\bibnamefont{Jarillo-Herrero}},
  \bibinfo{journal}{Nature} \textbf{\bibinfo{volume}{556}},
  \bibinfo{pages}{43–50} (\bibinfo{year}{2018}), ISSN
  \bibinfo{issn}{1476-4687},
  \urlprefix\url{http://dx.doi.org/10.1038/nature26160}.

\bibitem[{\citenamefont{Lu et~al.}(2019)\citenamefont{Lu, Stepanov, Yang, Xie,
  Aamir, Das, Urgell, Watanabe, Taniguchi, Zhang
  et~al.}}]{lu2019superconductors}
\bibinfo{author}{\bibfnamefont{X.}~\bibnamefont{Lu}},
  \bibinfo{author}{\bibfnamefont{P.}~\bibnamefont{Stepanov}},
  \bibinfo{author}{\bibfnamefont{W.}~\bibnamefont{Yang}},
  \bibinfo{author}{\bibfnamefont{M.}~\bibnamefont{Xie}},
  \bibinfo{author}{\bibfnamefont{M.~A.} \bibnamefont{Aamir}},
  \bibinfo{author}{\bibfnamefont{I.}~\bibnamefont{Das}},
  \bibinfo{author}{\bibfnamefont{C.}~\bibnamefont{Urgell}},
  \bibinfo{author}{\bibfnamefont{K.}~\bibnamefont{Watanabe}},
  \bibinfo{author}{\bibfnamefont{T.}~\bibnamefont{Taniguchi}},
  \bibinfo{author}{\bibfnamefont{G.}~\bibnamefont{Zhang}},
  \bibnamefont{et~al.}, \bibinfo{journal}{Nature}
  \textbf{\bibinfo{volume}{574}}, \bibinfo{pages}{653} (\bibinfo{year}{2019}).

\bibitem[{\citenamefont{Andrei and MacDonald}(2020)}]{andrei2020graphene}
\bibinfo{author}{\bibfnamefont{E.~Y.} \bibnamefont{Andrei}} \bibnamefont{and}
  \bibinfo{author}{\bibfnamefont{A.~H.} \bibnamefont{MacDonald}},
  \bibinfo{journal}{Nature materials} \textbf{\bibinfo{volume}{19}},
  \bibinfo{pages}{1265} (\bibinfo{year}{2020}).

\bibitem[{\citenamefont{Saito et~al.}(2020)\citenamefont{Saito, Ge, Watanabe,
  Taniguchi, and Young}}]{saito2020independent}
\bibinfo{author}{\bibfnamefont{Y.}~\bibnamefont{Saito}},
  \bibinfo{author}{\bibfnamefont{J.}~\bibnamefont{Ge}},
  \bibinfo{author}{\bibfnamefont{K.}~\bibnamefont{Watanabe}},
  \bibinfo{author}{\bibfnamefont{T.}~\bibnamefont{Taniguchi}},
  \bibnamefont{and} \bibinfo{author}{\bibfnamefont{A.~F.} \bibnamefont{Young}},
  \bibinfo{journal}{Nature Physics} \textbf{\bibinfo{volume}{16}},
  \bibinfo{pages}{926} (\bibinfo{year}{2020}).

\bibitem[{\citenamefont{Polshyn et~al.}(2019)\citenamefont{Polshyn, Yankowitz,
  Chen, Zhang, Watanabe, Taniguchi, Dean, and Young}}]{polshyn2019large}
\bibinfo{author}{\bibfnamefont{H.}~\bibnamefont{Polshyn}},
  \bibinfo{author}{\bibfnamefont{M.}~\bibnamefont{Yankowitz}},
  \bibinfo{author}{\bibfnamefont{S.}~\bibnamefont{Chen}},
  \bibinfo{author}{\bibfnamefont{Y.}~\bibnamefont{Zhang}},
  \bibinfo{author}{\bibfnamefont{K.}~\bibnamefont{Watanabe}},
  \bibinfo{author}{\bibfnamefont{T.}~\bibnamefont{Taniguchi}},
  \bibinfo{author}{\bibfnamefont{C.~R.} \bibnamefont{Dean}}, \bibnamefont{and}
  \bibinfo{author}{\bibfnamefont{A.~F.} \bibnamefont{Young}},
  \bibinfo{journal}{Nature Physics} \textbf{\bibinfo{volume}{15}},
  \bibinfo{pages}{1011} (\bibinfo{year}{2019}).

\bibitem[{\citenamefont{Oh et~al.}(2021)\citenamefont{Oh, Nuckolls, Wong, Lee,
  Liu, Watanabe, Taniguchi, and Yazdani}}]{Oh2021}
\bibinfo{author}{\bibfnamefont{M.}~\bibnamefont{Oh}},
  \bibinfo{author}{\bibfnamefont{K.~P.} \bibnamefont{Nuckolls}},
  \bibinfo{author}{\bibfnamefont{D.}~\bibnamefont{Wong}},
  \bibinfo{author}{\bibfnamefont{R.~L.} \bibnamefont{Lee}},
  \bibinfo{author}{\bibfnamefont{X.}~\bibnamefont{Liu}},
  \bibinfo{author}{\bibfnamefont{K.}~\bibnamefont{Watanabe}},
  \bibinfo{author}{\bibfnamefont{T.}~\bibnamefont{Taniguchi}},
  \bibnamefont{and} \bibinfo{author}{\bibfnamefont{A.}~\bibnamefont{Yazdani}},
  \bibinfo{journal}{Nature} \textbf{\bibinfo{volume}{600}},
  \bibinfo{pages}{240} (\bibinfo{year}{2021}),
  \urlprefix\url{https://doi.org/10.1038%2Fs41586-021-04121-x}.

\bibitem[{\citenamefont{Cao et~al.}(2021)\citenamefont{Cao, Rodan-Legrain,
  Park, Yuan, Watanabe, Taniguchi, Fernandes, Fu, and
  Jarillo-Herrero}}]{Cao_2021}
\bibinfo{author}{\bibfnamefont{Y.}~\bibnamefont{Cao}},
  \bibinfo{author}{\bibfnamefont{D.}~\bibnamefont{Rodan-Legrain}},
  \bibinfo{author}{\bibfnamefont{J.~M.} \bibnamefont{Park}},
  \bibinfo{author}{\bibfnamefont{N.~F.~Q.} \bibnamefont{Yuan}},
  \bibinfo{author}{\bibfnamefont{K.}~\bibnamefont{Watanabe}},
  \bibinfo{author}{\bibfnamefont{T.}~\bibnamefont{Taniguchi}},
  \bibinfo{author}{\bibfnamefont{R.~M.} \bibnamefont{Fernandes}},
  \bibinfo{author}{\bibfnamefont{L.}~\bibnamefont{Fu}}, \bibnamefont{and}
  \bibinfo{author}{\bibfnamefont{P.}~\bibnamefont{Jarillo-Herrero}},
  \bibinfo{journal}{Science} \textbf{\bibinfo{volume}{372}},
  \bibinfo{pages}{264} (\bibinfo{year}{2021}),
  \urlprefix\url{https://doi.org/10.1126%2Fscience.abc2836}.

\bibitem[{\citenamefont{Jaoui et~al.}(2022)\citenamefont{Jaoui, Das,
  Di~Battista, D{\'\i}ez-M{\'e}rida, Lu, Watanabe, Taniguchi, Ishizuka,
  Levitov, and Efetov}}]{jaoui2022quantum}
\bibinfo{author}{\bibfnamefont{A.}~\bibnamefont{Jaoui}},
  \bibinfo{author}{\bibfnamefont{I.}~\bibnamefont{Das}},
  \bibinfo{author}{\bibfnamefont{G.}~\bibnamefont{Di~Battista}},
  \bibinfo{author}{\bibfnamefont{J.}~\bibnamefont{D{\'\i}ez-M{\'e}rida}},
  \bibinfo{author}{\bibfnamefont{X.}~\bibnamefont{Lu}},
  \bibinfo{author}{\bibfnamefont{K.}~\bibnamefont{Watanabe}},
  \bibinfo{author}{\bibfnamefont{T.}~\bibnamefont{Taniguchi}},
  \bibinfo{author}{\bibfnamefont{H.}~\bibnamefont{Ishizuka}},
  \bibinfo{author}{\bibfnamefont{L.}~\bibnamefont{Levitov}}, \bibnamefont{and}
  \bibinfo{author}{\bibfnamefont{D.~K.} \bibnamefont{Efetov}},
  \bibinfo{journal}{Nature Physics} pp. \bibinfo{pages}{1--6}
  (\bibinfo{year}{2022}).

\bibitem[{\citenamefont{You and Vishwanath}(2019)}]{you2019superconductivity}
\bibinfo{author}{\bibfnamefont{Y.-Z.} \bibnamefont{You}} \bibnamefont{and}
  \bibinfo{author}{\bibfnamefont{A.}~\bibnamefont{Vishwanath}},
  \bibinfo{journal}{npj Quantum Materials} \textbf{\bibinfo{volume}{4}},
  \bibinfo{pages}{1} (\bibinfo{year}{2019}).

\bibitem[{\citenamefont{Kozii et~al.}(2019)\citenamefont{Kozii, Isobe,
  Venderbos, and Fu}}]{Kozii2019}
\bibinfo{author}{\bibfnamefont{V.}~\bibnamefont{Kozii}},
  \bibinfo{author}{\bibfnamefont{H.}~\bibnamefont{Isobe}},
  \bibinfo{author}{\bibfnamefont{J.~W.~F.} \bibnamefont{Venderbos}},
  \bibnamefont{and} \bibinfo{author}{\bibfnamefont{L.}~\bibnamefont{Fu}},
  \bibinfo{journal}{Phys. Rev. B} \textbf{\bibinfo{volume}{99}},
  \bibinfo{pages}{144507} (\bibinfo{year}{2019}),
  \urlprefix\url{https://link.aps.org/doi/10.1103/PhysRevB.99.144507}.

\bibitem[{\citenamefont{Chichinadze et~al.}(2020)\citenamefont{Chichinadze,
  Classen, and Chubukov}}]{Chichinadze2020}
\bibinfo{author}{\bibfnamefont{D.~V.} \bibnamefont{Chichinadze}},
  \bibinfo{author}{\bibfnamefont{L.}~\bibnamefont{Classen}}, \bibnamefont{and}
  \bibinfo{author}{\bibfnamefont{A.~V.} \bibnamefont{Chubukov}},
  \bibinfo{journal}{Phys. Rev. B} \textbf{\bibinfo{volume}{101}},
  \bibinfo{pages}{224513} (\bibinfo{year}{2020}),
  \urlprefix\url{https://link.aps.org/doi/10.1103/PhysRevB.101.224513}.

\bibitem[{\citenamefont{Khalaf et~al.}(2021)\citenamefont{Khalaf, Chatterjee,
  Bultinck, Zaletel, and Vishwanath}}]{khalaf2021charged}
\bibinfo{author}{\bibfnamefont{E.}~\bibnamefont{Khalaf}},
  \bibinfo{author}{\bibfnamefont{S.}~\bibnamefont{Chatterjee}},
  \bibinfo{author}{\bibfnamefont{N.}~\bibnamefont{Bultinck}},
  \bibinfo{author}{\bibfnamefont{M.~P.} \bibnamefont{Zaletel}},
  \bibnamefont{and}
  \bibinfo{author}{\bibfnamefont{A.}~\bibnamefont{Vishwanath}},
  \bibinfo{journal}{Science advances} \textbf{\bibinfo{volume}{7}},
  \bibinfo{pages}{eabf5299} (\bibinfo{year}{2021}).

\bibitem[{\citenamefont{Saito et~al.}(2019)\citenamefont{Saito, Ge, Watanabe,
  Taniguchi, and Young}}]{saito2019decoupling}
\bibinfo{author}{\bibfnamefont{Y.}~\bibnamefont{Saito}},
  \bibinfo{author}{\bibfnamefont{J.}~\bibnamefont{Ge}},
  \bibinfo{author}{\bibfnamefont{K.}~\bibnamefont{Watanabe}},
  \bibinfo{author}{\bibfnamefont{T.}~\bibnamefont{Taniguchi}},
  \bibnamefont{and} \bibinfo{author}{\bibfnamefont{A.~F.} \bibnamefont{Young}},
  \bibinfo{journal}{arXiv preprint arXiv:1911.13302}  (\bibinfo{year}{2019}).

\bibitem[{\citenamefont{Stepanov et~al.}(2020)\citenamefont{Stepanov, Das, Lu,
  Fahimniya, Watanabe, Taniguchi, Koppens, Lischner, Levitov, and
  Efetov}}]{stepanov2020untying}
\bibinfo{author}{\bibfnamefont{P.}~\bibnamefont{Stepanov}},
  \bibinfo{author}{\bibfnamefont{I.}~\bibnamefont{Das}},
  \bibinfo{author}{\bibfnamefont{X.}~\bibnamefont{Lu}},
  \bibinfo{author}{\bibfnamefont{A.}~\bibnamefont{Fahimniya}},
  \bibinfo{author}{\bibfnamefont{K.}~\bibnamefont{Watanabe}},
  \bibinfo{author}{\bibfnamefont{T.}~\bibnamefont{Taniguchi}},
  \bibinfo{author}{\bibfnamefont{F.~H.} \bibnamefont{Koppens}},
  \bibinfo{author}{\bibfnamefont{J.}~\bibnamefont{Lischner}},
  \bibinfo{author}{\bibfnamefont{L.}~\bibnamefont{Levitov}}, \bibnamefont{and}
  \bibinfo{author}{\bibfnamefont{D.~K.} \bibnamefont{Efetov}},
  \bibinfo{journal}{Nature} \textbf{\bibinfo{volume}{583}},
  \bibinfo{pages}{375} (\bibinfo{year}{2020}).

\bibitem[{\citenamefont{Wu et~al.}(2018)\citenamefont{Wu, MacDonald, and
  Martin}}]{Wu2018}
\bibinfo{author}{\bibfnamefont{F.}~\bibnamefont{Wu}},
  \bibinfo{author}{\bibfnamefont{A.~H.} \bibnamefont{MacDonald}},
  \bibnamefont{and} \bibinfo{author}{\bibfnamefont{I.}~\bibnamefont{Martin}},
  \bibinfo{journal}{Phys. Rev. Lett.} \textbf{\bibinfo{volume}{121}},
  \bibinfo{pages}{257001} (\bibinfo{year}{2018}),
  \urlprefix\url{https://link.aps.org/doi/10.1103/PhysRevLett.121.257001}.

\bibitem[{\citenamefont{Lian et~al.}(2019)\citenamefont{Lian, Wang, and
  Bernevig}}]{Lian2019}
\bibinfo{author}{\bibfnamefont{B.}~\bibnamefont{Lian}},
  \bibinfo{author}{\bibfnamefont{Z.}~\bibnamefont{Wang}}, \bibnamefont{and}
  \bibinfo{author}{\bibfnamefont{B.~A.} \bibnamefont{Bernevig}},
  \bibinfo{journal}{Phys. Rev. Lett.} \textbf{\bibinfo{volume}{122}},
  \bibinfo{pages}{257002} (\bibinfo{year}{2019}),
  \urlprefix\url{https://link.aps.org/doi/10.1103/PhysRevLett.122.257002}.

\bibitem[{\citenamefont{Wu et~al.}(2019)\citenamefont{Wu, Hwang, and
  Das~Sarma}}]{Wu2019}
\bibinfo{author}{\bibfnamefont{F.}~\bibnamefont{Wu}},
  \bibinfo{author}{\bibfnamefont{E.}~\bibnamefont{Hwang}}, \bibnamefont{and}
  \bibinfo{author}{\bibfnamefont{S.}~\bibnamefont{Das~Sarma}},
  \bibinfo{journal}{Phys. Rev. B} \textbf{\bibinfo{volume}{99}},
  \bibinfo{pages}{165112} (\bibinfo{year}{2019}),
  \urlprefix\url{https://link.aps.org/doi/10.1103/PhysRevB.99.165112}.

\bibitem[{\citenamefont{Liu et~al.}(2021)\citenamefont{Liu, Wang, Watanabe,
  Taniguchi, Vafek, and Li}}]{liu2021tuning}
\bibinfo{author}{\bibfnamefont{X.}~\bibnamefont{Liu}},
  \bibinfo{author}{\bibfnamefont{Z.}~\bibnamefont{Wang}},
  \bibinfo{author}{\bibfnamefont{K.}~\bibnamefont{Watanabe}},
  \bibinfo{author}{\bibfnamefont{T.}~\bibnamefont{Taniguchi}},
  \bibinfo{author}{\bibfnamefont{O.}~\bibnamefont{Vafek}}, \bibnamefont{and}
  \bibinfo{author}{\bibfnamefont{J.}~\bibnamefont{Li}},
  \bibinfo{journal}{Science} \textbf{\bibinfo{volume}{371}},
  \bibinfo{pages}{1261} (\bibinfo{year}{2021}).

\bibitem[{\citenamefont{Tarnopolsky et~al.}(2019)\citenamefont{Tarnopolsky,
  Kruchkov, and Vishwanath}}]{Tarnopolsky2019}
\bibinfo{author}{\bibfnamefont{G.}~\bibnamefont{Tarnopolsky}},
  \bibinfo{author}{\bibfnamefont{A.~J.} \bibnamefont{Kruchkov}},
  \bibnamefont{and}
  \bibinfo{author}{\bibfnamefont{A.}~\bibnamefont{Vishwanath}},
  \bibinfo{journal}{Phys. Rev. Lett.} \textbf{\bibinfo{volume}{122}},
  \bibinfo{pages}{106405} (\bibinfo{year}{2019}),
  \urlprefix\url{https://link.aps.org/doi/10.1103/PhysRevLett.122.106405}.

\bibitem[{\citenamefont{Zhou et~al.}(2021{\natexlab{a}})\citenamefont{Zhou,
  Xie, Ghazaryan, Holder, Ehrets, Spanton, Taniguchi, Watanabe, Berg, Serbyn
  et~al.}}]{zhou2021RTG}
\bibinfo{author}{\bibfnamefont{H.}~\bibnamefont{Zhou}},
  \bibinfo{author}{\bibfnamefont{T.}~\bibnamefont{Xie}},
  \bibinfo{author}{\bibfnamefont{A.}~\bibnamefont{Ghazaryan}},
  \bibinfo{author}{\bibfnamefont{T.}~\bibnamefont{Holder}},
  \bibinfo{author}{\bibfnamefont{J.~R.} \bibnamefont{Ehrets}},
  \bibinfo{author}{\bibfnamefont{E.~M.} \bibnamefont{Spanton}},
  \bibinfo{author}{\bibfnamefont{T.}~\bibnamefont{Taniguchi}},
  \bibinfo{author}{\bibfnamefont{K.}~\bibnamefont{Watanabe}},
  \bibinfo{author}{\bibfnamefont{E.}~\bibnamefont{Berg}},
  \bibinfo{author}{\bibfnamefont{M.}~\bibnamefont{Serbyn}},
  \bibnamefont{et~al.}, \bibinfo{journal}{Nature}
  \textbf{\bibinfo{volume}{598}}, \bibinfo{pages}{429–433}
  (\bibinfo{year}{2021}{\natexlab{a}}), ISSN \bibinfo{issn}{1476-4687},
  \urlprefix\url{http://dx.doi.org/10.1038/s41586-021-03938-w}.

\bibitem[{\citenamefont{Zhou et~al.}(2021{\natexlab{b}})\citenamefont{Zhou,
  Xie, Taniguchi, Watanabe, and Young}}]{zhou2021superconductivity}
\bibinfo{author}{\bibfnamefont{H.}~\bibnamefont{Zhou}},
  \bibinfo{author}{\bibfnamefont{T.}~\bibnamefont{Xie}},
  \bibinfo{author}{\bibfnamefont{T.}~\bibnamefont{Taniguchi}},
  \bibinfo{author}{\bibfnamefont{K.}~\bibnamefont{Watanabe}}, \bibnamefont{and}
  \bibinfo{author}{\bibfnamefont{A.~F.} \bibnamefont{Young}},
  \bibinfo{journal}{Nature} \textbf{\bibinfo{volume}{598}},
  \bibinfo{pages}{434} (\bibinfo{year}{2021}{\natexlab{b}}).

\bibitem[{\citenamefont{Zhou et~al.}(2021{\natexlab{c}})\citenamefont{Zhou,
  Saito, Cohen, Huynh, Patterson, Yang, Taniguchi, Watanabe, and
  Young}}]{zhou2021BLG}
\bibinfo{author}{\bibfnamefont{H.}~\bibnamefont{Zhou}},
  \bibinfo{author}{\bibfnamefont{Y.}~\bibnamefont{Saito}},
  \bibinfo{author}{\bibfnamefont{L.}~\bibnamefont{Cohen}},
  \bibinfo{author}{\bibfnamefont{W.}~\bibnamefont{Huynh}},
  \bibinfo{author}{\bibfnamefont{C.~L.} \bibnamefont{Patterson}},
  \bibinfo{author}{\bibfnamefont{F.}~\bibnamefont{Yang}},
  \bibinfo{author}{\bibfnamefont{T.}~\bibnamefont{Taniguchi}},
  \bibinfo{author}{\bibfnamefont{K.}~\bibnamefont{Watanabe}}, \bibnamefont{and}
  \bibinfo{author}{\bibfnamefont{A.~F.} \bibnamefont{Young}},
  \bibinfo{journal}{arXiv preprint arXiv:2110.11317}
  (\bibinfo{year}{2021}{\natexlab{c}}).

\bibitem[{\citenamefont{de~la Barrera et~al.}(2021)\citenamefont{de~la Barrera,
  Aronson, Zheng, Watanabe, Taniguchi, Ma, Jarillo-Herrero, and
  Ashoori}}]{de2021cascade}
\bibinfo{author}{\bibfnamefont{S.~C.} \bibnamefont{de~la Barrera}},
  \bibinfo{author}{\bibfnamefont{S.}~\bibnamefont{Aronson}},
  \bibinfo{author}{\bibfnamefont{Z.}~\bibnamefont{Zheng}},
  \bibinfo{author}{\bibfnamefont{K.}~\bibnamefont{Watanabe}},
  \bibinfo{author}{\bibfnamefont{T.}~\bibnamefont{Taniguchi}},
  \bibinfo{author}{\bibfnamefont{Q.}~\bibnamefont{Ma}},
  \bibinfo{author}{\bibfnamefont{P.}~\bibnamefont{Jarillo-Herrero}},
  \bibnamefont{and} \bibinfo{author}{\bibfnamefont{R.}~\bibnamefont{Ashoori}},
  \bibinfo{journal}{arXiv preprint arXiv:2110.13907}  (\bibinfo{year}{2021}).

\bibitem[{\citenamefont{McCann and Koshino}(2013)}]{McCann_2013}
\bibinfo{author}{\bibfnamefont{E.}~\bibnamefont{McCann}} \bibnamefont{and}
  \bibinfo{author}{\bibfnamefont{M.}~\bibnamefont{Koshino}},
  \bibinfo{journal}{Reports on Progress in Physics}
  \textbf{\bibinfo{volume}{76}}, \bibinfo{pages}{056503}
  (\bibinfo{year}{2013}),
  \urlprefix\url{https://doi.org/10.1088/0034-4885/76/5/056503}.

\bibitem[{\citenamefont{McCann and Fal'ko}(2006)}]{McCann2006Landau}
\bibinfo{author}{\bibfnamefont{E.}~\bibnamefont{McCann}} \bibnamefont{and}
  \bibinfo{author}{\bibfnamefont{V.~I.} \bibnamefont{Fal'ko}},
  \bibinfo{journal}{Phys. Rev. Lett.} \textbf{\bibinfo{volume}{96}},
  \bibinfo{pages}{086805} (\bibinfo{year}{2006}),
  \urlprefix\url{https://link.aps.org/doi/10.1103/PhysRevLett.96.086805}.

\bibitem[{\citenamefont{Nandkishore and
  Levitov}(2010{\natexlab{a}})}]{Nandkishore2010dynamical}
\bibinfo{author}{\bibfnamefont{R.}~\bibnamefont{Nandkishore}} \bibnamefont{and}
  \bibinfo{author}{\bibfnamefont{L.}~\bibnamefont{Levitov}},
  \bibinfo{journal}{Phys. Rev. Lett.} \textbf{\bibinfo{volume}{104}},
  \bibinfo{pages}{156803} (\bibinfo{year}{2010}{\natexlab{a}}),
  \urlprefix\url{https://link.aps.org/doi/10.1103/PhysRevLett.104.156803}.

\bibitem[{\citenamefont{Nandkishore and
  Levitov}(2010{\natexlab{b}})}]{Nandkishore2010quantum}
\bibinfo{author}{\bibfnamefont{R.}~\bibnamefont{Nandkishore}} \bibnamefont{and}
  \bibinfo{author}{\bibfnamefont{L.}~\bibnamefont{Levitov}},
  \bibinfo{journal}{Phys. Rev. B} \textbf{\bibinfo{volume}{82}},
  \bibinfo{pages}{115124} (\bibinfo{year}{2010}{\natexlab{b}}),
  \urlprefix\url{https://link.aps.org/doi/10.1103/PhysRevB.82.115124}.

\bibitem[{\citenamefont{Vafek and Yang}(2010)}]{Vafek2010}
\bibinfo{author}{\bibfnamefont{O.}~\bibnamefont{Vafek}} \bibnamefont{and}
  \bibinfo{author}{\bibfnamefont{K.}~\bibnamefont{Yang}},
  \bibinfo{journal}{Phys. Rev. B} \textbf{\bibinfo{volume}{81}},
  \bibinfo{pages}{041401} (\bibinfo{year}{2010}),
  \urlprefix\url{https://link.aps.org/doi/10.1103/PhysRevB.81.041401}.

\bibitem[{\citenamefont{Jung et~al.}(2011)\citenamefont{Jung, Zhang, and
  MacDonald}}]{Jung2011lattice}
\bibinfo{author}{\bibfnamefont{J.}~\bibnamefont{Jung}},
  \bibinfo{author}{\bibfnamefont{F.}~\bibnamefont{Zhang}}, \bibnamefont{and}
  \bibinfo{author}{\bibfnamefont{A.~H.} \bibnamefont{MacDonald}},
  \bibinfo{journal}{Phys. Rev. B} \textbf{\bibinfo{volume}{83}},
  \bibinfo{pages}{115408} (\bibinfo{year}{2011}),
  \urlprefix\url{https://link.aps.org/doi/10.1103/PhysRevB.83.115408}.

\bibitem[{\citenamefont{MacDonald et~al.}(2012)\citenamefont{MacDonald, Jung,
  and Zhang}}]{macdonald2012pseudospin}
\bibinfo{author}{\bibfnamefont{A.~H.} \bibnamefont{MacDonald}},
  \bibinfo{author}{\bibfnamefont{J.}~\bibnamefont{Jung}}, \bibnamefont{and}
  \bibinfo{author}{\bibfnamefont{F.}~\bibnamefont{Zhang}},
  \bibinfo{journal}{Physica Scripta} \textbf{\bibinfo{volume}{2012}},
  \bibinfo{pages}{014012} (\bibinfo{year}{2012}).

\bibitem[{\citenamefont{Zhang and MacDonald}(2012)}]{Zhang2012Distinguishing}
\bibinfo{author}{\bibfnamefont{F.}~\bibnamefont{Zhang}} \bibnamefont{and}
  \bibinfo{author}{\bibfnamefont{A.~H.} \bibnamefont{MacDonald}},
  \bibinfo{journal}{Phys. Rev. Lett.} \textbf{\bibinfo{volume}{108}},
  \bibinfo{pages}{186804} (\bibinfo{year}{2012}),
  \urlprefix\url{https://link.aps.org/doi/10.1103/PhysRevLett.108.186804}.

\bibitem[{\citenamefont{Cvetkovic et~al.}(2012)\citenamefont{Cvetkovic,
  Throckmorton, and Vafek}}]{Cvetkovic2012}
\bibinfo{author}{\bibfnamefont{V.}~\bibnamefont{Cvetkovic}},
  \bibinfo{author}{\bibfnamefont{R.~E.} \bibnamefont{Throckmorton}},
  \bibnamefont{and} \bibinfo{author}{\bibfnamefont{O.}~\bibnamefont{Vafek}},
  \bibinfo{journal}{Phys. Rev. B} \textbf{\bibinfo{volume}{86}},
  \bibinfo{pages}{075467} (\bibinfo{year}{2012}),
  \urlprefix\url{https://link.aps.org/doi/10.1103/PhysRevB.86.075467}.

\bibitem[{\citenamefont{Throckmorton and Das~Sarma}(2014)}]{Throckmorton2014}
\bibinfo{author}{\bibfnamefont{R.~E.} \bibnamefont{Throckmorton}}
  \bibnamefont{and}
  \bibinfo{author}{\bibfnamefont{S.}~\bibnamefont{Das~Sarma}},
  \bibinfo{journal}{Phys. Rev. B} \textbf{\bibinfo{volume}{90}},
  \bibinfo{pages}{205407} (\bibinfo{year}{2014}),
  \urlprefix\url{https://link.aps.org/doi/10.1103/PhysRevB.90.205407}.

\bibitem[{\citenamefont{Min et~al.}(2008)\citenamefont{Min, Borghi, Polini, and
  MacDonald}}]{Min2008}
\bibinfo{author}{\bibfnamefont{H.}~\bibnamefont{Min}},
  \bibinfo{author}{\bibfnamefont{G.}~\bibnamefont{Borghi}},
  \bibinfo{author}{\bibfnamefont{M.}~\bibnamefont{Polini}}, \bibnamefont{and}
  \bibinfo{author}{\bibfnamefont{A.~H.} \bibnamefont{MacDonald}},
  \bibinfo{journal}{Phys. Rev. B} \textbf{\bibinfo{volume}{77}},
  \bibinfo{pages}{041407} (\bibinfo{year}{2008}),
  \urlprefix\url{https://link.aps.org/doi/10.1103/PhysRevB.77.041407}.

\bibitem[{\citenamefont{Nilsson et~al.}(2006)\citenamefont{Nilsson,
  Castro~Neto, Peres, and Guinea}}]{Nilsson2006}
\bibinfo{author}{\bibfnamefont{J.}~\bibnamefont{Nilsson}},
  \bibinfo{author}{\bibfnamefont{A.~H.} \bibnamefont{Castro~Neto}},
  \bibinfo{author}{\bibfnamefont{N.~M.~R.} \bibnamefont{Peres}},
  \bibnamefont{and} \bibinfo{author}{\bibfnamefont{F.}~\bibnamefont{Guinea}},
  \bibinfo{journal}{Phys. Rev. B} \textbf{\bibinfo{volume}{73}},
  \bibinfo{pages}{214418} (\bibinfo{year}{2006}),
  \urlprefix\url{https://link.aps.org/doi/10.1103/PhysRevB.73.214418}.

\bibitem[{\citenamefont{Martin et~al.}(2010)\citenamefont{Martin, Feldman,
  Weitz, Allen, and Yacoby}}]{martin2010local}
\bibinfo{author}{\bibfnamefont{J.}~\bibnamefont{Martin}},
  \bibinfo{author}{\bibfnamefont{B.~E.} \bibnamefont{Feldman}},
  \bibinfo{author}{\bibfnamefont{R.~T.} \bibnamefont{Weitz}},
  \bibinfo{author}{\bibfnamefont{M.~T.} \bibnamefont{Allen}}, \bibnamefont{and}
  \bibinfo{author}{\bibfnamefont{A.}~\bibnamefont{Yacoby}},
  \bibinfo{journal}{Phys. Rev. Lett.} \textbf{\bibinfo{volume}{105}},
  \bibinfo{pages}{256806} (\bibinfo{year}{2010}),
  \urlprefix\url{https://link.aps.org/doi/10.1103/PhysRevLett.105.256806}.

\bibitem[{\citenamefont{Weitz et~al.}(2010)\citenamefont{Weitz, Allen, Feldman,
  Martin, and Yacoby}}]{weitz2010broken}
\bibinfo{author}{\bibfnamefont{R.~T.} \bibnamefont{Weitz}},
  \bibinfo{author}{\bibfnamefont{M.~T.} \bibnamefont{Allen}},
  \bibinfo{author}{\bibfnamefont{B.~E.} \bibnamefont{Feldman}},
  \bibinfo{author}{\bibfnamefont{J.}~\bibnamefont{Martin}}, \bibnamefont{and}
  \bibinfo{author}{\bibfnamefont{A.}~\bibnamefont{Yacoby}},
  \bibinfo{journal}{Science} \textbf{\bibinfo{volume}{330}},
  \bibinfo{pages}{812} (\bibinfo{year}{2010}).

\bibitem[{\citenamefont{Mayorov et~al.}(2011)\citenamefont{Mayorov, Elias,
  Mucha-Kruczynski, Gorbachev, Tudorovskiy, Zhukov, Morozov, Katsnelson, null
  null, Geim et~al.}}]{Mayorov2011}
\bibinfo{author}{\bibfnamefont{A.~S.} \bibnamefont{Mayorov}},
  \bibinfo{author}{\bibfnamefont{D.~C.} \bibnamefont{Elias}},
  \bibinfo{author}{\bibfnamefont{M.}~\bibnamefont{Mucha-Kruczynski}},
  \bibinfo{author}{\bibfnamefont{R.~V.} \bibnamefont{Gorbachev}},
  \bibinfo{author}{\bibfnamefont{T.}~\bibnamefont{Tudorovskiy}},
  \bibinfo{author}{\bibfnamefont{A.}~\bibnamefont{Zhukov}},
  \bibinfo{author}{\bibfnamefont{S.~V.} \bibnamefont{Morozov}},
  \bibinfo{author}{\bibfnamefont{M.~I.} \bibnamefont{Katsnelson}},
  \bibinfo{author}{\bibnamefont{null null}},
  \bibinfo{author}{\bibfnamefont{A.~K.} \bibnamefont{Geim}},
  \bibnamefont{et~al.}, \bibinfo{journal}{Science}
  \textbf{\bibinfo{volume}{333}}, \bibinfo{pages}{860} (\bibinfo{year}{2011}),
  \eprint{https://www.science.org/doi/pdf/10.1126/science.1208683},
  \urlprefix\url{https://www.science.org/doi/abs/10.1126/science.1208683}.

\bibitem[{\citenamefont{Veligura et~al.}(2012)\citenamefont{Veligura, van
  Elferen, Tombros, Maan, Zeitler, and van Wees}}]{Veligura2012}
\bibinfo{author}{\bibfnamefont{A.}~\bibnamefont{Veligura}},
  \bibinfo{author}{\bibfnamefont{H.~J.} \bibnamefont{van Elferen}},
  \bibinfo{author}{\bibfnamefont{N.}~\bibnamefont{Tombros}},
  \bibinfo{author}{\bibfnamefont{J.~C.} \bibnamefont{Maan}},
  \bibinfo{author}{\bibfnamefont{U.}~\bibnamefont{Zeitler}}, \bibnamefont{and}
  \bibinfo{author}{\bibfnamefont{B.~J.} \bibnamefont{van Wees}},
  \bibinfo{journal}{Phys. Rev. B} \textbf{\bibinfo{volume}{85}},
  \bibinfo{pages}{155412} (\bibinfo{year}{2012}),
  \urlprefix\url{https://link.aps.org/doi/10.1103/PhysRevB.85.155412}.

\bibitem[{\citenamefont{Velasco et~al.}(2012)\citenamefont{Velasco, Jing, Bao,
  Lee, Kratz, Aji, Bockrath, Lau, Varma, Stillwell
  et~al.}}]{velasco2012transport}
\bibinfo{author}{\bibfnamefont{J.}~\bibnamefont{Velasco}},
  \bibinfo{author}{\bibfnamefont{L.}~\bibnamefont{Jing}},
  \bibinfo{author}{\bibfnamefont{W.}~\bibnamefont{Bao}},
  \bibinfo{author}{\bibfnamefont{Y.}~\bibnamefont{Lee}},
  \bibinfo{author}{\bibfnamefont{P.}~\bibnamefont{Kratz}},
  \bibinfo{author}{\bibfnamefont{V.}~\bibnamefont{Aji}},
  \bibinfo{author}{\bibfnamefont{M.}~\bibnamefont{Bockrath}},
  \bibinfo{author}{\bibfnamefont{C.}~\bibnamefont{Lau}},
  \bibinfo{author}{\bibfnamefont{C.}~\bibnamefont{Varma}},
  \bibinfo{author}{\bibfnamefont{R.}~\bibnamefont{Stillwell}},
  \bibnamefont{et~al.}, \bibinfo{journal}{Nature nanotechnology}
  \textbf{\bibinfo{volume}{7}}, \bibinfo{pages}{156} (\bibinfo{year}{2012}).

\bibitem[{\citenamefont{Bao et~al.}(2012)\citenamefont{Bao, Velasco, Zhang,
  Jing, Standley, Smirnov, Bockrath, MacDonald, and Lau}}]{Bao2012}
\bibinfo{author}{\bibfnamefont{W.}~\bibnamefont{Bao}},
  \bibinfo{author}{\bibfnamefont{J.}~\bibnamefont{Velasco}},
  \bibinfo{author}{\bibfnamefont{F.}~\bibnamefont{Zhang}},
  \bibinfo{author}{\bibfnamefont{L.}~\bibnamefont{Jing}},
  \bibinfo{author}{\bibfnamefont{B.}~\bibnamefont{Standley}},
  \bibinfo{author}{\bibfnamefont{D.}~\bibnamefont{Smirnov}},
  \bibinfo{author}{\bibfnamefont{M.}~\bibnamefont{Bockrath}},
  \bibinfo{author}{\bibfnamefont{A.~H.} \bibnamefont{MacDonald}},
  \bibnamefont{and} \bibinfo{author}{\bibfnamefont{C.~N.} \bibnamefont{Lau}},
  \bibinfo{journal}{Proceedings of the National Academy of Sciences}
  \textbf{\bibinfo{volume}{109}}, \bibinfo{pages}{10802}
  (\bibinfo{year}{2012}),
  \eprint{https://www.pnas.org/doi/pdf/10.1073/pnas.1205978109},
  \urlprefix\url{https://www.pnas.org/doi/abs/10.1073/pnas.1205978109}.

\bibitem[{\citenamefont{Freitag et~al.}(2012)\citenamefont{Freitag, Trbovic,
  Weiss, and Sch\"onenberger}}]{Freitag2012}
\bibinfo{author}{\bibfnamefont{F.}~\bibnamefont{Freitag}},
  \bibinfo{author}{\bibfnamefont{J.}~\bibnamefont{Trbovic}},
  \bibinfo{author}{\bibfnamefont{M.}~\bibnamefont{Weiss}}, \bibnamefont{and}
  \bibinfo{author}{\bibfnamefont{C.}~\bibnamefont{Sch\"onenberger}},
  \bibinfo{journal}{Phys. Rev. Lett.} \textbf{\bibinfo{volume}{108}},
  \bibinfo{pages}{076602} (\bibinfo{year}{2012}),
  \urlprefix\url{https://link.aps.org/doi/10.1103/PhysRevLett.108.076602}.

\bibitem[{\citenamefont{Seiler et~al.}(2021)\citenamefont{Seiler, Geisenhof,
  Winterer, Watanabe, Taniguchi, Xu, Zhang, and Weitz}}]{Seiler2021}
\bibinfo{author}{\bibfnamefont{A.~M.} \bibnamefont{Seiler}},
  \bibinfo{author}{\bibfnamefont{F.~R.} \bibnamefont{Geisenhof}},
  \bibinfo{author}{\bibfnamefont{F.}~\bibnamefont{Winterer}},
  \bibinfo{author}{\bibfnamefont{K.}~\bibnamefont{Watanabe}},
  \bibinfo{author}{\bibfnamefont{T.}~\bibnamefont{Taniguchi}},
  \bibinfo{author}{\bibfnamefont{T.}~\bibnamefont{Xu}},
  \bibinfo{author}{\bibfnamefont{F.}~\bibnamefont{Zhang}}, \bibnamefont{and}
  \bibinfo{author}{\bibfnamefont{R.~T.} \bibnamefont{Weitz}},
  \bibinfo{journal}{arXiv preprint arXiv:2111.06413}  (\bibinfo{year}{2021}).

\bibitem[{\citenamefont{McCann}(2006)}]{McCann2006Asymmetry}
\bibinfo{author}{\bibfnamefont{E.}~\bibnamefont{McCann}},
  \bibinfo{journal}{Phys. Rev. B} \textbf{\bibinfo{volume}{74}},
  \bibinfo{pages}{161403} (\bibinfo{year}{2006}),
  \urlprefix\url{https://link.aps.org/doi/10.1103/PhysRevB.74.161403}.

\bibitem[{\citenamefont{Saito et~al.}(2021)\citenamefont{Saito, Yang, Ge, Liu,
  Taniguchi, Watanabe, Li, Berg, and Young}}]{saito2021isospin}
\bibinfo{author}{\bibfnamefont{Y.}~\bibnamefont{Saito}},
  \bibinfo{author}{\bibfnamefont{F.}~\bibnamefont{Yang}},
  \bibinfo{author}{\bibfnamefont{J.}~\bibnamefont{Ge}},
  \bibinfo{author}{\bibfnamefont{X.}~\bibnamefont{Liu}},
  \bibinfo{author}{\bibfnamefont{T.}~\bibnamefont{Taniguchi}},
  \bibinfo{author}{\bibfnamefont{K.}~\bibnamefont{Watanabe}},
  \bibinfo{author}{\bibfnamefont{J.}~\bibnamefont{Li}},
  \bibinfo{author}{\bibfnamefont{E.}~\bibnamefont{Berg}}, \bibnamefont{and}
  \bibinfo{author}{\bibfnamefont{A.~F.} \bibnamefont{Young}},
  \bibinfo{journal}{Nature} \textbf{\bibinfo{volume}{592}},
  \bibinfo{pages}{220} (\bibinfo{year}{2021}).

\bibitem[{\citenamefont{Zondiner et~al.}(2020)\citenamefont{Zondiner, Rozen,
  Rodan-Legrain, Cao, Queiroz, Taniguchi, Watanabe, Oreg, von Oppen, Stern
  et~al.}}]{zondiner2020cascade}
\bibinfo{author}{\bibfnamefont{U.}~\bibnamefont{Zondiner}},
  \bibinfo{author}{\bibfnamefont{A.}~\bibnamefont{Rozen}},
  \bibinfo{author}{\bibfnamefont{D.}~\bibnamefont{Rodan-Legrain}},
  \bibinfo{author}{\bibfnamefont{Y.}~\bibnamefont{Cao}},
  \bibinfo{author}{\bibfnamefont{R.}~\bibnamefont{Queiroz}},
  \bibinfo{author}{\bibfnamefont{T.}~\bibnamefont{Taniguchi}},
  \bibinfo{author}{\bibfnamefont{K.}~\bibnamefont{Watanabe}},
  \bibinfo{author}{\bibfnamefont{Y.}~\bibnamefont{Oreg}},
  \bibinfo{author}{\bibfnamefont{F.}~\bibnamefont{von Oppen}},
  \bibinfo{author}{\bibfnamefont{A.}~\bibnamefont{Stern}},
  \bibnamefont{et~al.}, \bibinfo{journal}{Nature}
  \textbf{\bibinfo{volume}{582}}, \bibinfo{pages}{203} (\bibinfo{year}{2020}).

\bibitem[{\citenamefont{Rozen et~al.}(2021)\citenamefont{Rozen, Park, Zondiner,
  Cao, Rodan-Legrain, Taniguchi, Watanabe, Oreg, Stern, Berg
  et~al.}}]{rozen2021entropic}
\bibinfo{author}{\bibfnamefont{A.}~\bibnamefont{Rozen}},
  \bibinfo{author}{\bibfnamefont{J.~M.} \bibnamefont{Park}},
  \bibinfo{author}{\bibfnamefont{U.}~\bibnamefont{Zondiner}},
  \bibinfo{author}{\bibfnamefont{Y.}~\bibnamefont{Cao}},
  \bibinfo{author}{\bibfnamefont{D.}~\bibnamefont{Rodan-Legrain}},
  \bibinfo{author}{\bibfnamefont{T.}~\bibnamefont{Taniguchi}},
  \bibinfo{author}{\bibfnamefont{K.}~\bibnamefont{Watanabe}},
  \bibinfo{author}{\bibfnamefont{Y.}~\bibnamefont{Oreg}},
  \bibinfo{author}{\bibfnamefont{A.}~\bibnamefont{Stern}},
  \bibinfo{author}{\bibfnamefont{E.}~\bibnamefont{Berg}}, \bibnamefont{et~al.},
  \bibinfo{journal}{Nature} \textbf{\bibinfo{volume}{592}},
  \bibinfo{pages}{214} (\bibinfo{year}{2021}).

\bibitem[{\citenamefont{Choi et~al.}(2021)\citenamefont{Choi, Kim, Peng,
  Thomson, Lewandowski, Polski, Zhang, Arora, Watanabe, Taniguchi
  et~al.}}]{choi2021correlation}
\bibinfo{author}{\bibfnamefont{Y.}~\bibnamefont{Choi}},
  \bibinfo{author}{\bibfnamefont{H.}~\bibnamefont{Kim}},
  \bibinfo{author}{\bibfnamefont{Y.}~\bibnamefont{Peng}},
  \bibinfo{author}{\bibfnamefont{A.}~\bibnamefont{Thomson}},
  \bibinfo{author}{\bibfnamefont{C.}~\bibnamefont{Lewandowski}},
  \bibinfo{author}{\bibfnamefont{R.}~\bibnamefont{Polski}},
  \bibinfo{author}{\bibfnamefont{Y.}~\bibnamefont{Zhang}},
  \bibinfo{author}{\bibfnamefont{H.~S.} \bibnamefont{Arora}},
  \bibinfo{author}{\bibfnamefont{K.}~\bibnamefont{Watanabe}},
  \bibinfo{author}{\bibfnamefont{T.}~\bibnamefont{Taniguchi}},
  \bibnamefont{et~al.}, \bibinfo{journal}{Nature}
  \textbf{\bibinfo{volume}{589}}, \bibinfo{pages}{536} (\bibinfo{year}{2021}).

\bibitem[{\citenamefont{Pierce et~al.}(2021)\citenamefont{Pierce, Xie, Park,
  Khalaf, Lee, Cao, Parker, Forrester, Chen, Watanabe
  et~al.}}]{pierce2021unconventional}
\bibinfo{author}{\bibfnamefont{A.~T.} \bibnamefont{Pierce}},
  \bibinfo{author}{\bibfnamefont{Y.}~\bibnamefont{Xie}},
  \bibinfo{author}{\bibfnamefont{J.~M.} \bibnamefont{Park}},
  \bibinfo{author}{\bibfnamefont{E.}~\bibnamefont{Khalaf}},
  \bibinfo{author}{\bibfnamefont{S.~H.} \bibnamefont{Lee}},
  \bibinfo{author}{\bibfnamefont{Y.}~\bibnamefont{Cao}},
  \bibinfo{author}{\bibfnamefont{D.~E.} \bibnamefont{Parker}},
  \bibinfo{author}{\bibfnamefont{P.~R.} \bibnamefont{Forrester}},
  \bibinfo{author}{\bibfnamefont{S.}~\bibnamefont{Chen}},
  \bibinfo{author}{\bibfnamefont{K.}~\bibnamefont{Watanabe}},
  \bibnamefont{et~al.}, \bibinfo{journal}{Nature Physics}
  \textbf{\bibinfo{volume}{17}}, \bibinfo{pages}{1210} (\bibinfo{year}{2021}).

\bibitem[{\citenamefont{Jung et~al.}(2015)\citenamefont{Jung, Polini, and
  MacDonald}}]{Jung2015}
\bibinfo{author}{\bibfnamefont{J.}~\bibnamefont{Jung}},
  \bibinfo{author}{\bibfnamefont{M.}~\bibnamefont{Polini}}, \bibnamefont{and}
  \bibinfo{author}{\bibfnamefont{A.~H.} \bibnamefont{MacDonald}},
  \bibinfo{journal}{Phys. Rev. B} \textbf{\bibinfo{volume}{91}},
  \bibinfo{pages}{155423} (\bibinfo{year}{2015}),
  \urlprefix\url{https://link.aps.org/doi/10.1103/PhysRevB.91.155423}.

\bibitem[{\citenamefont{Dong et~al.}(2021)\citenamefont{Dong, Davydova,
  Ogunnaike, and Levitov}}]{dong2021}
\bibinfo{author}{\bibfnamefont{Z.}~\bibnamefont{Dong}},
  \bibinfo{author}{\bibfnamefont{M.}~\bibnamefont{Davydova}},
  \bibinfo{author}{\bibfnamefont{M.}~\bibnamefont{Ogunnaike}},
  \bibnamefont{and} \bibinfo{author}{\bibfnamefont{L.}~\bibnamefont{Levitov}},
  \bibinfo{journal}{arXiv preprint arXiv:2110.15254}  (\bibinfo{year}{2021}).

\bibitem[{\citenamefont{Ghazaryan et~al.}(2021)\citenamefont{Ghazaryan, Holder,
  Serbyn, and Berg}}]{Ghazaryan2021}
\bibinfo{author}{\bibfnamefont{A.}~\bibnamefont{Ghazaryan}},
  \bibinfo{author}{\bibfnamefont{T.}~\bibnamefont{Holder}},
  \bibinfo{author}{\bibfnamefont{M.}~\bibnamefont{Serbyn}}, \bibnamefont{and}
  \bibinfo{author}{\bibfnamefont{E.}~\bibnamefont{Berg}},
  \bibinfo{journal}{Phys. Rev. Lett.} \textbf{\bibinfo{volume}{127}},
  \bibinfo{pages}{247001} (\bibinfo{year}{2021}),
  \urlprefix\url{https://link.aps.org/doi/10.1103/PhysRevLett.127.247001}.

\bibitem[{\citenamefont{Chatterjee et~al.}(2021)\citenamefont{Chatterjee, Wang,
  Berg, and Zaletel}}]{chatterjee2021inter}
\bibinfo{author}{\bibfnamefont{S.}~\bibnamefont{Chatterjee}},
  \bibinfo{author}{\bibfnamefont{T.}~\bibnamefont{Wang}},
  \bibinfo{author}{\bibfnamefont{E.}~\bibnamefont{Berg}}, \bibnamefont{and}
  \bibinfo{author}{\bibfnamefont{M.~P.} \bibnamefont{Zaletel}},
  \bibinfo{journal}{arXiv preprint arXiv:2109.00002}  (\bibinfo{year}{2021}).

\bibitem[{\citenamefont{You and Vishwanath}(2022)}]{You2022}
\bibinfo{author}{\bibfnamefont{Y.-Z.} \bibnamefont{You}} \bibnamefont{and}
  \bibinfo{author}{\bibfnamefont{A.}~\bibnamefont{Vishwanath}},
  \bibinfo{journal}{Phys. Rev. B} \textbf{\bibinfo{volume}{105}},
  \bibinfo{pages}{134524} (\bibinfo{year}{2022}),
  \urlprefix\url{https://link.aps.org/doi/10.1103/PhysRevB.105.134524}.

\bibitem[{\citenamefont{Chou et~al.}(2022)\citenamefont{Chou, Wu, Sau, and
  Das~Sarma}}]{Chou2022}
\bibinfo{author}{\bibfnamefont{Y.-Z.} \bibnamefont{Chou}},
  \bibinfo{author}{\bibfnamefont{F.}~\bibnamefont{Wu}},
  \bibinfo{author}{\bibfnamefont{J.~D.} \bibnamefont{Sau}}, \bibnamefont{and}
  \bibinfo{author}{\bibfnamefont{S.}~\bibnamefont{Das~Sarma}},
  \bibinfo{journal}{Phys. Rev. B} \textbf{\bibinfo{volume}{105}},
  \bibinfo{pages}{L100503} (\bibinfo{year}{2022}),
  \urlprefix\url{https://link.aps.org/doi/10.1103/PhysRevB.105.L100503}.

\bibitem[{\citenamefont{Klein et~al.}(2020)\citenamefont{Klein, Chubukov,
  Schattner, and Berg}}]{Klein2020}
\bibinfo{author}{\bibfnamefont{A.}~\bibnamefont{Klein}},
  \bibinfo{author}{\bibfnamefont{A.~V.} \bibnamefont{Chubukov}},
  \bibinfo{author}{\bibfnamefont{Y.}~\bibnamefont{Schattner}},
  \bibnamefont{and} \bibinfo{author}{\bibfnamefont{E.}~\bibnamefont{Berg}},
  \bibinfo{journal}{Phys. Rev. X} \textbf{\bibinfo{volume}{10}},
  \bibinfo{pages}{031053} (\bibinfo{year}{2020}),
  \urlprefix\url{https://link.aps.org/doi/10.1103/PhysRevX.10.031053}.

\bibitem[{\citenamefont{Oganesyan et~al.}(2001)\citenamefont{Oganesyan,
  Kivelson, and Fradkin}}]{Oganesyan2001}
\bibinfo{author}{\bibfnamefont{V.}~\bibnamefont{Oganesyan}},
  \bibinfo{author}{\bibfnamefont{S.~A.} \bibnamefont{Kivelson}},
  \bibnamefont{and} \bibinfo{author}{\bibfnamefont{E.}~\bibnamefont{Fradkin}},
  \bibinfo{journal}{Phys. Rev. B} \textbf{\bibinfo{volume}{64}},
  \bibinfo{pages}{195109} (\bibinfo{year}{2001}),
  \urlprefix\url{https://link.aps.org/doi/10.1103/PhysRevB.64.195109}.

\bibitem[{\citenamefont{Lederer et~al.}(2015)\citenamefont{Lederer, Schattner,
  Berg, and Kivelson}}]{Lederer2015}
\bibinfo{author}{\bibfnamefont{S.}~\bibnamefont{Lederer}},
  \bibinfo{author}{\bibfnamefont{Y.}~\bibnamefont{Schattner}},
  \bibinfo{author}{\bibfnamefont{E.}~\bibnamefont{Berg}}, \bibnamefont{and}
  \bibinfo{author}{\bibfnamefont{S.~A.} \bibnamefont{Kivelson}},
  \bibinfo{journal}{Phys. Rev. Lett.} \textbf{\bibinfo{volume}{114}},
  \bibinfo{pages}{097001} (\bibinfo{year}{2015}),
  \urlprefix\url{https://link.aps.org/doi/10.1103/PhysRevLett.114.097001}.

\bibitem[{\citenamefont{Tremblay}(2013)}]{Tremblay2013}
\bibinfo{author}{\bibfnamefont{A.-M.} \bibnamefont{Tremblay}},
  \bibinfo{journal}{Emergent Phenomena in Correlated Matter Modeling and
  Simulation} \textbf{\bibinfo{volume}{3}} (\bibinfo{year}{2013}).

\bibitem[{\citenamefont{Chubukov et~al.}(2020)\citenamefont{Chubukov, Abanov,
  Wang, and Wu}}]{Chubukov_2020a}
\bibinfo{author}{\bibfnamefont{A.~V.} \bibnamefont{Chubukov}},
  \bibinfo{author}{\bibfnamefont{A.}~\bibnamefont{Abanov}},
  \bibinfo{author}{\bibfnamefont{Y.}~\bibnamefont{Wang}}, \bibnamefont{and}
  \bibinfo{author}{\bibfnamefont{Y.-M.} \bibnamefont{Wu}},
  \bibinfo{journal}{Annals of Physics} p. \bibinfo{pages}{168142}
  (\bibinfo{year}{2020}), ISSN \bibinfo{issn}{0003-4916},
  \urlprefix\url{http://www.sciencedirect.com/science/article/pii/S0003491620300750}.

\bibitem[{\citenamefont{Berg et~al.}(2019)\citenamefont{Berg, Lederer,
  Schattner, and Trebst}}]{berg_4}
\bibinfo{author}{\bibfnamefont{E.}~\bibnamefont{Berg}},
  \bibinfo{author}{\bibfnamefont{S.}~\bibnamefont{Lederer}},
  \bibinfo{author}{\bibfnamefont{Y.}~\bibnamefont{Schattner}},
  \bibnamefont{and} \bibinfo{author}{\bibfnamefont{S.}~\bibnamefont{Trebst}},
  \bibinfo{journal}{Annual Review of Condensed Matter Physics}
  \textbf{\bibinfo{volume}{10}}, \bibinfo{pages}{63} (\bibinfo{year}{2019}),
  \urlprefix\url{https://doi.org/10.1146/annurev-conmatphys-031218-013339}.

\bibitem[{\citenamefont{Lee}(2018)}]{sslee_2018}
\bibinfo{author}{\bibfnamefont{S.-S.} \bibnamefont{Lee}},
  \bibinfo{journal}{Annual Review of Condensed Matter Physics}
  \textbf{\bibinfo{volume}{9}}, \bibinfo{pages}{227} (\bibinfo{year}{2018}),
  \urlprefix\url{https://doi.org/10.1146/annurev-conmatphys-031016-025531}.

\bibitem[{\citenamefont{Efetov et~al.}(2013)\citenamefont{Efetov, Meier, and
  Pepin}}]{efetov2013}
\bibinfo{author}{\bibfnamefont{K.~B.} \bibnamefont{Efetov}},
  \bibinfo{author}{\bibfnamefont{H.}~\bibnamefont{Meier}}, \bibnamefont{and}
  \bibinfo{author}{\bibfnamefont{C.}~\bibnamefont{Pepin}},
  \bibinfo{journal}{Nature Physics} \textbf{\bibinfo{volume}{9}},
  \bibinfo{pages}{442} (\bibinfo{year}{2013}).

\bibitem[{\citenamefont{Mazin et~al.}(2008)\citenamefont{Mazin, Singh,
  Johannes, and Du}}]{Mazin2008unconventional}
\bibinfo{author}{\bibfnamefont{I.~I.} \bibnamefont{Mazin}},
  \bibinfo{author}{\bibfnamefont{D.~J.} \bibnamefont{Singh}},
  \bibinfo{author}{\bibfnamefont{M.~D.} \bibnamefont{Johannes}},
  \bibnamefont{and} \bibinfo{author}{\bibfnamefont{M.~H.} \bibnamefont{Du}},
  \bibinfo{journal}{Phys. Rev. Lett.} \textbf{\bibinfo{volume}{101}},
  \bibinfo{pages}{057003} (\bibinfo{year}{2008}),
  \urlprefix\url{https://link.aps.org/doi/10.1103/PhysRevLett.101.057003}.

\bibitem[{\citenamefont{Mazin and Schmalian}(2009)}]{Mazin2009pairing}
\bibinfo{author}{\bibfnamefont{I.}~\bibnamefont{Mazin}} \bibnamefont{and}
  \bibinfo{author}{\bibfnamefont{J.}~\bibnamefont{Schmalian}},
  \bibinfo{journal}{Physica C: Superconductivity}
  \textbf{\bibinfo{volume}{469}}, \bibinfo{pages}{614} (\bibinfo{year}{2009}),
  ISSN \bibinfo{issn}{0921-4534}, \bibinfo{note}{superconductivity in
  Iron-Pnictides},
  \urlprefix\url{https://www.sciencedirect.com/science/article/pii/S0921453409001002}.

\bibitem[{\citenamefont{Scalapino et~al.}(1966)\citenamefont{Scalapino,
  Schrieffer, and Wilkins}}]{Scalapino1966}
\bibinfo{author}{\bibfnamefont{D.~J.} \bibnamefont{Scalapino}},
  \bibinfo{author}{\bibfnamefont{J.~R.} \bibnamefont{Schrieffer}},
  \bibnamefont{and} \bibinfo{author}{\bibfnamefont{J.~W.}
  \bibnamefont{Wilkins}}, \bibinfo{journal}{Phys. Rev.}
  \textbf{\bibinfo{volume}{148}}, \bibinfo{pages}{263} (\bibinfo{year}{1966}),
  \urlprefix\url{https://link.aps.org/doi/10.1103/PhysRev.148.263}.

\bibitem[{SM()}]{SM}
\emph{\bibinfo{title}{Supplemental material [url will be inserted by
  publisher].}}

\bibitem[{\citenamefont{Morel and Anderson}(1962)}]{Morel1962}
\bibinfo{author}{\bibfnamefont{P.}~\bibnamefont{Morel}} \bibnamefont{and}
  \bibinfo{author}{\bibfnamefont{P.~W.} \bibnamefont{Anderson}},
  \bibinfo{journal}{Phys. Rev.} \textbf{\bibinfo{volume}{125}},
  \bibinfo{pages}{1263} (\bibinfo{year}{1962}),
  \urlprefix\url{https://link.aps.org/doi/10.1103/PhysRev.125.1263}.

\bibitem[{\citenamefont{McMillan}(1968)}]{McMillan1968}
\bibinfo{author}{\bibfnamefont{W.~L.} \bibnamefont{McMillan}},
  \bibinfo{journal}{Phys. Rev.} \textbf{\bibinfo{volume}{167}},
  \bibinfo{pages}{331} (\bibinfo{year}{1968}),
  \urlprefix\url{https://link.aps.org/doi/10.1103/PhysRev.167.331}.

\bibitem[{\citenamefont{Bogoljubov et~al.}(1958)\citenamefont{Bogoljubov,
  Tolmachov, and Širkov}}]{Bogoljubov1958}
\bibinfo{author}{\bibfnamefont{N.~N.} \bibnamefont{Bogoljubov}},
  \bibinfo{author}{\bibfnamefont{V.~V.} \bibnamefont{Tolmachov}},
  \bibnamefont{and} \bibinfo{author}{\bibfnamefont{D.~V.}
  \bibnamefont{Širkov}}, \bibinfo{journal}{Fortschritte der Physik}
  \textbf{\bibinfo{volume}{6}}, \bibinfo{pages}{605} (\bibinfo{year}{1958}),
  \eprint{https://onlinelibrary.wiley.com/doi/pdf/10.1002/prop.19580061102},
  \urlprefix\url{https://onlinelibrary.wiley.com/doi/abs/10.1002/prop.19580061102}.

\bibitem[{\citenamefont{Coleman}(2015)}]{Coleman2015}
\bibinfo{author}{\bibfnamefont{P.}~\bibnamefont{Coleman}},
  \emph{\bibinfo{title}{Introduction to Many-Body Physics}}
  (\bibinfo{publisher}{Cambridge University Press}, \bibinfo{year}{2015}), ISBN
  \bibinfo{isbn}{9781139020916},
  \urlprefix\url{https://doi.org/10.1017/cbo9781139020916}.

\bibitem[{\citenamefont{Chubukov et~al.}(2019)\citenamefont{Chubukov,
  Prokof'ev, and Svistunov}}]{prokofiev}
\bibinfo{author}{\bibfnamefont{A.}~\bibnamefont{Chubukov}},
  \bibinfo{author}{\bibfnamefont{N.~V.} \bibnamefont{Prokof'ev}},
  \bibnamefont{and} \bibinfo{author}{\bibfnamefont{B.~V.}
  \bibnamefont{Svistunov}}, \bibinfo{journal}{Phys. Rev. B}
  \textbf{\bibinfo{volume}{100}}, \bibinfo{pages}{064513}
  (\bibinfo{year}{2019}),
  \urlprefix\url{https://link.aps.org/doi/10.1103/PhysRevB.100.064513}.

\bibitem[{\citenamefont{Kohn and Luttinger}(1965)}]{Kohn1965}
\bibinfo{author}{\bibfnamefont{W.}~\bibnamefont{Kohn}} \bibnamefont{and}
  \bibinfo{author}{\bibfnamefont{J.~M.} \bibnamefont{Luttinger}},
  \bibinfo{journal}{Phys. Rev. Lett.} \textbf{\bibinfo{volume}{15}},
  \bibinfo{pages}{524} (\bibinfo{year}{1965}),
  \urlprefix\url{https://link.aps.org/doi/10.1103/PhysRevLett.15.524}.

\bibitem[{\citenamefont{Bonesteel et~al.}(1996)\citenamefont{Bonesteel,
  McDonald, and Nayak}}]{Bonesteel1996}
\bibinfo{author}{\bibfnamefont{N.~E.} \bibnamefont{Bonesteel}},
  \bibinfo{author}{\bibfnamefont{I.~A.} \bibnamefont{McDonald}},
  \bibnamefont{and} \bibinfo{author}{\bibfnamefont{C.}~\bibnamefont{Nayak}},
  \bibinfo{journal}{Phys. Rev. Lett.} \textbf{\bibinfo{volume}{77}},
  \bibinfo{pages}{3009} (\bibinfo{year}{1996}),
  \urlprefix\url{https://link.aps.org/doi/10.1103/PhysRevLett.77.3009}.

\bibitem[{\citenamefont{Ruhman and Lee}(2016)}]{ruhman}
\bibinfo{author}{\bibfnamefont{J.}~\bibnamefont{Ruhman}} \bibnamefont{and}
  \bibinfo{author}{\bibfnamefont{P.~A.} \bibnamefont{Lee}},
  \bibinfo{journal}{Phys. Rev. B} \textbf{\bibinfo{volume}{94}},
  \bibinfo{pages}{224515} (\bibinfo{year}{2016}),
  \urlprefix\url{https://link.aps.org/doi/10.1103/PhysRevB.94.224515}.

\bibitem[{\citenamefont{Pimenov and Chubukov}(2022)}]{pimenov}
\bibinfo{author}{\bibfnamefont{D.}~\bibnamefont{Pimenov}} \bibnamefont{and}
  \bibinfo{author}{\bibfnamefont{A.~V.} \bibnamefont{Chubukov}},
  \bibinfo{journal}{npj Quantum Materials} \textbf{\bibinfo{volume}{7}},
  \bibinfo{pages}{1} (\bibinfo{year}{2022}).

\bibitem[{\citenamefont{Maslov et~al.}(2011)\citenamefont{Maslov, Yudson, and
  Chubukov}}]{Maslov2011}
\bibinfo{author}{\bibfnamefont{D.~L.} \bibnamefont{Maslov}},
  \bibinfo{author}{\bibfnamefont{V.~I.} \bibnamefont{Yudson}},
  \bibnamefont{and} \bibinfo{author}{\bibfnamefont{A.~V.}
  \bibnamefont{Chubukov}}, \bibinfo{journal}{Phys. Rev. Lett.}
  \textbf{\bibinfo{volume}{106}}, \bibinfo{pages}{106403}
  (\bibinfo{year}{2011}),
  \urlprefix\url{https://link.aps.org/doi/10.1103/PhysRevLett.106.106403}.

\bibitem[{\citenamefont{Chubukov and W\"olfle}(2014)}]{wolfle}
\bibinfo{author}{\bibfnamefont{A.~V.} \bibnamefont{Chubukov}} \bibnamefont{and}
  \bibinfo{author}{\bibfnamefont{P.}~\bibnamefont{W\"olfle}},
  \bibinfo{journal}{Phys. Rev. B} \textbf{\bibinfo{volume}{89}},
  \bibinfo{pages}{045108} (\bibinfo{year}{2014}),
  \urlprefix\url{https://link.aps.org/doi/10.1103/PhysRevB.89.045108}.

\end{thebibliography}

\begin{appendix}
	
	\section{Diagrammatic derivation of the effective interaction.}\label{SM:diagrams}
	
	It is instructive to analyze how the diagrammatic series, which we use to obtain the effective pairing interaction $\Gamma_{ss} (\nu, q)$, emerge in order-by-order expansion in $V_0$.
	
	By general rules, the vertex function $\Gamma_{ss'} (\nu, q)$  is the fully dressed antisymmetrized interaction, irreducible in a particular channel. For pairing we need irreducible interaction  between fermions with momenta $(k,-k; p, -p)$.
	The specifics of our case is that (i) pairing involves one fermion near $K$ and one near $K'$,
	and (ii) the scattering from $K$ to $K'$ is weak in graphene and can be neglected.  One can check that in this situation
	the contributions to $\Gamma_{ss} (\nu, q)$ from antisymmetrization vanish, and we can restrict with just dressed interaction $V_0$.  As we are interested in pairing in a $B$ field, we focus on the vertex function with equal spin projections, $\Gamma_{ss} (\nu, q)$.
	
\begin{figure}[t]
		\centering
		\includegraphics[width=0.15\textwidth]{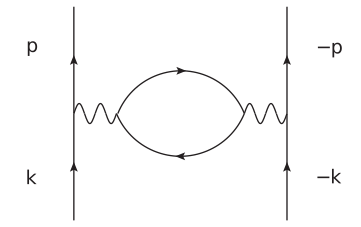}
		\includegraphics[width=0.15\textwidth]{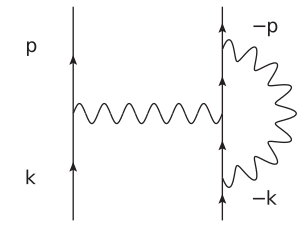}
		\includegraphics[width=0.15\textwidth]{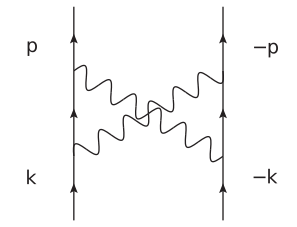}
		\caption{Two types of second-order diagrams that are enhanced near the onset of $q=0$ Stoner order.
			a) bubble diagram b) ``wine glass" diagram c) ``exchange" diagram
		}\label{fig:second order diagram}
	\end{figure}

\begin{figure}[t]
		\centering
		\includegraphics[width=0.49\textwidth]{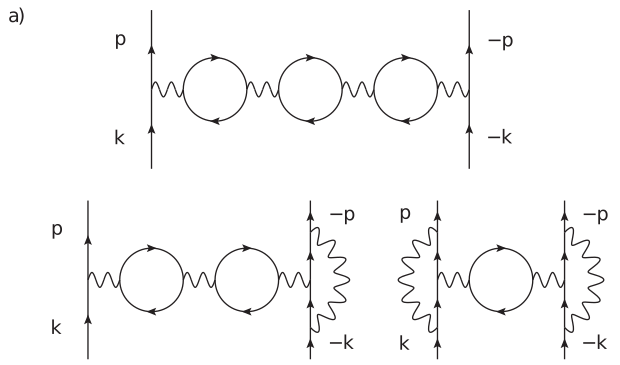}
		\includegraphics[width=0.49\textwidth]{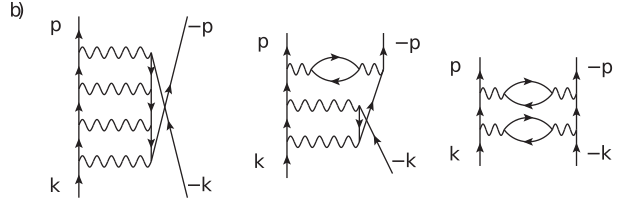}
		\caption{Diagrams at 3-loop order.
			a) diagrams with 3 bubbles b) diagrams with 0,1 or 2, which are subleading near the Stoner transition.}
\label{fig:3-loop diagram}
	\end{figure}

	To first order in $V_0$ the vertex function is just a constant $V_0$. To second-order we have three topologically different sets of diagrams, shown in Fig.\ref{fig:second order diagram} (they are often called KL diagrams).  The diagrams of the first two sets (bubble and "wine glass" diagrams) are expressed via the polarization bubble at momenta $q= ({\bf k}-{\bf p})$, the last "exchange" diagram is expressed via the polarization bubble at momenta $q_{+} = {\bf k} + {\bf p}$.  We are primarily interested in the dressed interaction at small momentum transfer as we expect that this interaction gets enhanced near the onset of $q=0$ isospin order.  Accordingly, at each order of expansion in $V_0$ we will only use
	diagrams, which contain polarization bubbles with $q$ and neglect diagrams, which contain bubbles with $q_{+}$
	(see Ref.\cite{wolfle} for similar consideration).
	The corresponding diagrams at 3-loop order are shown in Fig.\ref{fig:3-loop diagram} They contain diagrams with zero, one, and two bubbles, and one diagram with interaction line inserted into the bubble.
	
	This structure suggests the way to sum up relevant diagrams by separating them into sub-classes with different number
	of bubbles, and within each subclass inserting all possible interaction lines into  the bubbles.  A small experimentation shows that this leads to diagrammatic series shown in Fig.\ref{fig:leading diagram}.
	Diagrams without bubbles sum up into $V_0 \gamma^2_s (\nu, q)$, where $\gamma_s (\nu, q)$ is given by a ladder series of $V_0 \Pi_s (\nu, q)$.  An insertion of a bare bubble brings the factor $-2 V_0 \left(\Pi_{\uparrow} (\nu, q)+ \Pi_{\downarrow} (\nu, q)\right)$, where $-1$ is due to a loop and the factor $2$ comes about because
	intermediate fermions can be near $K$  or near $K'$ no matter where external fermions are.   The insertion of ladder series of interactions into each bubble further multiples $\Pi_{s} (\nu, q)$ by $\gamma_{s} (\nu, q)$. The end result is Eq. (\ref{eq:H_eff1}) in the main text.
	
	\medskip

	\section{Self-energy}
In this section, we elaborate on the effect of self-energy correction, and show that it is unimportant.
Here, we focus on the case of $\nu_1=0$. The self-energy for spin-down electrons is given by
\bea
\Sigma_{\downarrow\downarrow}(\omega) &&= \int \frac{d\nu}{2\pi} \int \frac{dq_{\perp}}{2\pi}  \frac{\overline\Gamma_{\downarrow\downarrow}(\nu,q)}{i(\omega+\nu) - v_F q_{\perp}}\\
 && = \frac{\lambda}{2}\int \frac{d\nu}{2\pi} \sgn(\nu + \omega)S(\nu/\nu_0)\\
 && = \lambda \nu_0 \int_0^{{\bar\omega}} dx S(x)
\eea
To see the relevant scale of $\omega$ in pairing problem, we look at the gap equation without self-energy correction \eqref{eq:gap equation_1}. We rewrite it as follows:
\be
\Delta(n) = - \frac{\lambda}{2}   \sum_{n'} \frac{\Delta(n')} {|2n'+1|}S\left(2\pi{\bar T}_c \lp n-n'\rp,0\right)
\label{eq:gap equation_1_SM}
\ee
where we have rewritten $\bar\omega = \pi (2n+1){\bar T_c}$, $\bar\omega' = \pi (2n'+1){\bar T_c}$. As shown by numerics in the main text, the critical temperature is $T_c \sim \frac{1}{2} \lambda \nu_0\times 10^{-2}$. For extremely large value of $\lambda\gg 10^2$ (i.e. extremely small  $\delta_\downarrow$), the relevant $n,n'\sim O(1)$.
Replacing $S(x)$ with its asymptotic form $\frac{1}{2x}$, we find numerically that the wavefunction stop changing sign at $n= 4$. The self-energy relevant for this pairing problem should be evaluated at $\omega=9\pi T_c$:
\be
\Sigma_{\downarrow\downarrow}(\omega\sim 9\pi T_c) \sim \lambda \nu_0 \ln(\lambda)
\ee
where logarithm comes from integrating $S(x)$ which scales as $\frac{1}{2x}$ at large $x$.
Then, we find the effective coupling scales with $\lambda$ as
\be
\tilde{\lambda} = \frac{\lambda}{1+\kappa \ln  \lambda}, \quad \kappa \sim \frac{100}{9\pi} = 3.5
\ee
The denominator is only marginally relevant at large $\lambda$, thus does not suppress the $T_c$ substantially.

For a not-so-large $\lambda$ value ($\lambda<10^2$), the relevant value of $n,n'$ the equation above is $n,n'\lesssim\nu_0/T_c =10^2/\lambda $.
The self-energy relevant for this pairing problem should be evaluated at $\omega \lesssim \nu_0$:
\be
\Sigma_{\downarrow\downarrow}(\omega\sim \nu_0) \sim 0.2 \lambda \nu_0
\ee
where the numerical factor comes from integrating $S(x)$ below the turning pint $x\sim 0.5$.
We find the effective coupling scales with $\lambda$ as
\be
\tilde{\lambda} = \frac{\lambda}{1+  0.2 \lambda}.
\ee
In this case, carrying out the simulation as Fig.\ref{fig:Tc} of main text, we get $T_c\sim 10^{-4}$, which is much smaller than the value in Fig.\ref{fig:Tc}, but is still of an acceptable orders of magnitude.

\end{appendix}

\end{document}